\documentclass[journal,draftclsnofoot,12pt,onecolumn]{IEEEtran}
\usepackage[cmex10]{amsmath}
\usepackage{xmpaper,lzqsymbol}
\usepackage{amssymb,latexsym}
\usepackage{graphicx,subfigure}
\usepackage{subfigure}
\usepackage{color,cite,times}
\usepackage{epstopdf}
\usepackage{multirow}
\usepackage{algorithm}
\usepackage{algorithmic}
\usepackage{booktabs}
\usepackage{bbm} 
\usepackage{subfigure}
\usepackage{url}
\usepackage[pdftex, bookmarksnumbered, bookmarksopen]{hyperref}

\newcommand{\nc}{\newcommand}
\nc{\bsm}{\boldsymbol}
\nc{\mbs}{\mathbbmss}

\begin{document}
\title{Frequency Reflection Modulation for Reconfigurable Intelligent Surface Aided OFDM Systems}

\author{Wenjing Yan, Xiaojun~Yuan,~\IEEEmembership{Senior~Member,~IEEE}, and Xuanyu~Cao,~\IEEEmembership{Senior~Member,~IEEE} 
\thanks{W. Yan was with the National Laboratory of Science and Technology on Communications, the University of Electronic Science and Technology of China, Chengdu, and is now with the Department of Electronic and Computer Engineering, The Hong Kong University of Science and Technology, Hong Kong (e-mail: wj.yan@connect.ust.hk).
X. Yuan is with the National Laboratory of Science and Technology on Communications, the University of Electronic Science and Technology of China, Chengdu,
China (e-mail: xjyuan@uestc.edu.cn). X. Cao is  with the Department of Electronic and Computer Engineering, The Hong Kong University of Science and Technology, Hong Kong (e-mail: eexcao@ust.hk). (The corresponding author is X. Yuan.)}
}

\maketitle
\begin{abstract}
    Reconfigurable intelligent surface (RIS) based reflection modulation (RM) has been considered as a promising information delivery mechanism, and has the potential to realize passive information transfer of a RIS without consuming any additional radio frequency chain and time/frequency/energy resource. The existing on-off RM (ORM) schemes are based on manipulating the ``on/off'' states of RIS reflection elements, which may lead to the degradation of RIS reflection efficiency. This paper proposes a frequency RM (FRM) method for RIS-aided OFDM systems. The FRM-OFDM scheme modulates the frequency of the incident electromagnetic waves, and the RIS information is embedded in the frequency-hopping states of RIS elements. Unlike the ORM-OFDM scheme, the FRM-OFDM scheme can achieve higher reflection efficiency, since the latter does not turn off any reflection element in RM. We show that, for the RIS phase shift optimization, the multiplicative multiple access channel in the FRM-OFDM system can be converted to an equivalent RIS-aided multiple-input multiple-output channel. Then, we propose an alternating optimization (AO) algorithm for sum rate maximization of the FRM-OFDM system. A low-complexity recursive AO algorithm is further developed to avoid direct channel matrix inversion in the AO algorithm with negligible performance degradation.
    In addition, we design a bilinear message passing (BMP) algorithm for the bilinear recovery of both the user symbols and the RIS data. Numerical simulations verify the efficiency of the designed optimization algorithms for system optimization and the BMP algorithm for signal detection, as well as the superiority of the proposed FRM-OFDM scheme over the existing ORM-OFDM scheme and the RIS-aided OFDM system. 
\end{abstract}

\section{Introduction}

Reconfigurable intelligent surfaces (RISs), as large electromagnetic metasurfaces, consist of numerous low-cost and nearly-passive reflection elements, each of which is able to independently induce a phase shift to incident electromagnetic (EM) waves \cite{cui2014coding,di2019smart}. By collaboratively designing these phase shifts, the reflection property of a RIS can be artificially manipulated to enhance the performance of wireless networks, such as expanding communication coverage, improving received signal-to-noise ratio (SNR), mitigating interference and/or eavesdropping, and so on \cite{wu2019Towards}. Compared to a multi-antenna relay, a prominent advantage of a RIS is that it reflects impinging EM waves in a passive manner and requires no additional time/frequency/energy resources or hardware cost for processing, re-generating, or re-transmitting signals. These appealing features have motivated a variety of emerging research directions on RIS, such as cascaded channel estimation \cite{he2019cascaded,taha2019enabling,wang2019channel,liu2020matrix}, passive beamforming (PB) \cite{wu2019intelligent,huang2019reconfigurable,cai2021hierarchical}, RIS-aided communications \cite{cai2020reconfigurable,li2020reconfigurable,yang2021ris,zhang2021airis}, information modulation techniques \cite{basar2020reconfigurable,yan2019passive,guo2020reflecting,lin2020reconfigurable,gopi2020intelligent}, and hardware implementations \cite{tang2020Path,dai2020reconfigurable}.

The need of RIS information transfer arises in various ways. First, the RIS in a communication system needs to report/upload handshake information during the link establishment process, and send out the confirmation when synchronizing with the transceiver for packet transmission. Second, the configuration and maintenance of a RIS-aided communication system also generate data at the RIS end for delivery \cite{yuan2021reconfigurable,guo2020reflecting}. For example, real-time monitoring of the RIS working environment (such as temperature, humidity, pressure, etc.) is necessary to avoid the impairment of RIS elements and to report failure if such impairment happens. 
RIS information transfer can be achieved by equipping a RIS with a dedicated radio frequency (RF) chain, but this compromises the “passive” nature of the RIS by consuming extra time/frequency resources and increases the hardware cost for RIS implementation. As such, it is desirable to appropriately design passive information transmission (PIT) at the RIS end, which is the focus of this paper.

Reflection modulation (RM) \cite{basar2020reconfigurable,yan2019passive,guo2020reflecting,lin2020reconfigurable,gopi2020intelligent} has been considered as a promising PIT mechanism for RIS-aided communication systems, in which multiple information streams are encoded in both the carriers emitted by transmitter(s) and the reflection patterns of the RIS. Compared with the dedicated RF chain approach to uploading RIS information, the RM mechanism possesses the following three main advantages: 1) No additional hardware cost and energy consumption for generating or retransmitting RF signals; 2) No consumption of extra time/frequency resource since the RIS data are delivered together with the active information transfer; 3) more importantly, RM has the potential to improve the multiplexing gain of the RIS-aided system. It is shown in \cite{cheng2021degree} that in a multiple-input multiple-output (MIMO) channel with $U$ transmit antennas and $M$ receive antennas, an RM-based RIS with $N$ reflective elements can improve the degrees-of-freedom of the system from $\min(U;M)$ to $\min(U + \frac{N}{2}- \frac{1}{2};N;M)$ in the absence of direct link.

Recent developments on the PIT of RIS are briefly described as follows. 
In \cite{basar2020reconfigurable}, the author proposed spatial modulation on the indices of receive antennas, which is realized by steering the reflected beam of RIS towards a particular receive antenna.  A beam-index modulation scheme is proposed in \cite{gopi2020intelligent} for RIS-aided millimeter wave systems based on a twin-RIS structure, where the RIS’s beam-pattern is randomized for carrying information. The authors in \cite{yan2019passive} proposed a joint PB and information transfer (PBIT) scheme by conducting spatial modulation on the RIS elements. In the PBIT scheme, the “on/off” state of each RIS element is randomly adjusted to represent an information bit, which is referred to as “on-off RM (ORM)”. In \cite{yan2020passive}, the same group of authors further extended the work to MIMO ORM systems, and studied the design of the RIS phase shifts and the bilinear receiver. The ORM scheme studied in \cite{yan2019passive} and \cite{yan2020passive} assumes that each RIS element is tuned on independently, which has a risk of link-outage as the number of activated elements fluctuates. To address this issue, the authors in \cite{lin2020reconfigurable} proposed a RIS-based reflection pattern modulation (RIS-RPM) scheme, in which the number of RIS elements switched on at any time instance is fixed. However, there are always a proportion of turned-off RIS elements in ORM schemes, which compromises the PB capability of a RIS.

In this paper, to avoid the loss of PB capability suffered by ORM, we propose a frequency RM (FRM) method for the RIS-aided orthogonal frequency division multiplexing (OFDM) system, where a single-antenna user transmits OFDM modulated signals to a multi-antenna base station (BS) via the help of a RIS. In the FRM-OFDM scheme, each RIS element modulates its incident EM waves via frequency hopping, and the RIS information is embedded in the frequency-hopping states of the RIS elements. The frequency hopping on a RIS element is realized by tuning the phase shift of the element continuously over time, where the feasibility of continuously changing RIS phase shifts has been experimentally verified in \cite{tang2020mimo}. The frequency-hopped signals are shifted to adjacent subcarriers (SCs), and are collected by the OFDM receiver to enhance the system performance. Unlike an ORM-OFDM system where the turned-off elements sacrifice PB capability, the FRM-OFDM scheme carries out full-on reflection and so is able to achieve a higher reflection efficiency.

We study the sum rate maximization of the FRM-OFDM system by optimizing the RIS phase shifts. 
Similar to ORM-OFDM, the simultaneous transmission of user and RIS information in the FRM-OFDM gives rise to a multiplicative multiple access channel (MMAC). Maximizing the sum rate of the MMAC is a challenging task. To simplify this problem, prior works \cite{yan2019passive,yan2020passive,lin2020reconfigurable} proposed to maximize a lower bound by ignoring the mutual information between the RIS data and the received signals. This paper takes the first attempt to directly cope with the sum rate maximization problem in the MMAC channel.  We first show that, with a Gaussian approximation on the distribution of the received signals, the MMAC channel of the FRM-OFDM system can be converted to an equivalent multiple-input multiple-output (MIMO) channel in the sense of RIS phase shift optimization. Then, we propose an alternating optimization (AO) algorithm for the sum-rate optimization by following the minimum mean-square error (MMSE) procedure.

The AO algorithm involves direct channel matrix inversion, which is computational involving especially for large-size FRM-OFDM systems. To avoid this problem, we further develop a low-complexity recursive AO (RAO) algorithm by decomposing the calculation of the sum rate in a recursive manner, which reduces the computational complexity of the AO algorithm by $\mathcal{O}(K^2)$ times, with $K$ being the number of SCs in the OFDM system.

Moreover, we study the receiver design of the FRM-OFDM system, which gives rise to a bilinear problem involving the joint detection of user symbols and RIS states. We design a message passing framework to address this problem, in which the user symbols are retrieved by forward and backward recursions after maximum ratio combining (MRC) on the received signals, and the RIS states are detected by the generalized approximate message passing (GAMP) algorithm \cite{rangan2011generalized}. 

We show by numerical simulations that the proposed FRM-OFDM systems achieves a much broader rate region than the ORM-OFDM system, since that former realizes RM by turning on all the RIS elements. Furthermore, we show that the FRM-OFDM system achieves a sharper rate slope against the signal-to-noise ratio (SNR) than the conventional RIS-aided OFDM system (in which the RIS is used solely for PB). This verifies the multiplexing gain of the RM reported in \cite{cheng2021degree}

The main contributions of this paper are summarized as follows.
\begin{itemize}
    \item  We propose an FRM-OFDM scheme for the joint PB and PIT of RIS with higher reflection efficiency than the existing ORM-OFDM scheme. 
    \item  We show that the MMAC channel of the FRM-OFDM system can be converted to an equivalent RIS-aided MIMO channel in the sense of RIS phase shift optimization, and propose an AO algorithm to maximize the sum rate of the FRM-OFDM system.  
    \item  A low-complexity RAO algorithm is further developed to ease the complexity caused by direct inversion over the MIMO channel matrix as in the AO algorithm with negligible performance degradation. 
    \item  A bilinear message passing (BMP) algorithm is designed to address the bilinear signal recovery of both the user symbols and the RIS states.  
 \end{itemize}

 {\it Organization:} The rest of this paper is organized as follows. Section II introduces the main idea of the proposed FRM-OFDM scheme, and provides mathematical models to characterize the system. Section III describes the RIS phase shift optimization for sum rate maximization. Section IV proposes the BMP algorithm for the receiver design. Numerical results are presented in Section V. Finally, we conclude this work in Section VI.

{\it Notation:}
For any matrix $\bsm{A}$, $\bsm{a}_i$ refers to the $i$th column of $\bsm{A}$, and $a_{ij}$ refers to the $(i,j)$th entry of $\bsm{A}$.
 $\bsm{a}_{[i:j]}$ takes the $i$th to the $j$th entries in $\bsm{a}$. $\bsm{A}_{\rm diag }^{M\times N}$ and  $\bsm{A}_{\rm diag-}^{M\times N}$ are respectively the block diagonal and block lower sub-diagonal matrices taken from $\bsm{A}$ with block size $M\times N$. $\bsm{A}_{\rm diag+}^{M\times N} = \left((\bsm{A}^{\rm T})_{\rm diag-}^{M\times N}\right)^{\rm T}$. $\mathbb{R}$ and $\mathbb{C}$ denote the real field and complex field, respectively; $\mathcal{S}$ denotes a set, and $|\mathcal{S}|$ represents the cardinality of $\mathcal{S}$. $|x|$ represents the absolute value of $x$; $\|\cdot\|_2$ represents the $\ell_2$-norm. The superscripts $(\cdot)^{\rm T}$, $(\cdot)^{\ast}$, $(\cdot)^{\rm H}$, $(\cdot)^{-1}$ respectively represent the  transpose, the conjugate, the conjugate transpose, and the inverse of a matrix. $\odot$,  $\otimes$,  and $\circledast$ represent the Hadamard product, the Kronecker product, and the cyclic convolution, respectively.
 $\mbs{E}(\cdot)$ and $\mbs{Var}(\cdot)$ represent the expectation and the variance, respectively.
 $\delta(\cdot)$  represents the Dirac delta function. $\textbf{1}_N$ (or $\textbf{0}_N$) represents the $N$-dimensional all-one (or all-zero) vector, and $\textbf{I}_N$ represents the $N$-dimensional identity matrix.
 For any integer $N$,  $\mathcal{I}_N$ denotes the set of integers from $1$ to $N$.
 $\mathcal{CN}(\cdot;\mu,\nu)$ represents a complex Gaussian distribution with mean $\mu$, covariance $\nu$, and relation zero.

\section{Frequency Reflection Modulation in RIS-Aided OFDM Systems} \label{System model}
\subsection{RIS-Aided OFDM Systems}
\begin{figure}[t]
    \centering
    \includegraphics[width=3 in]{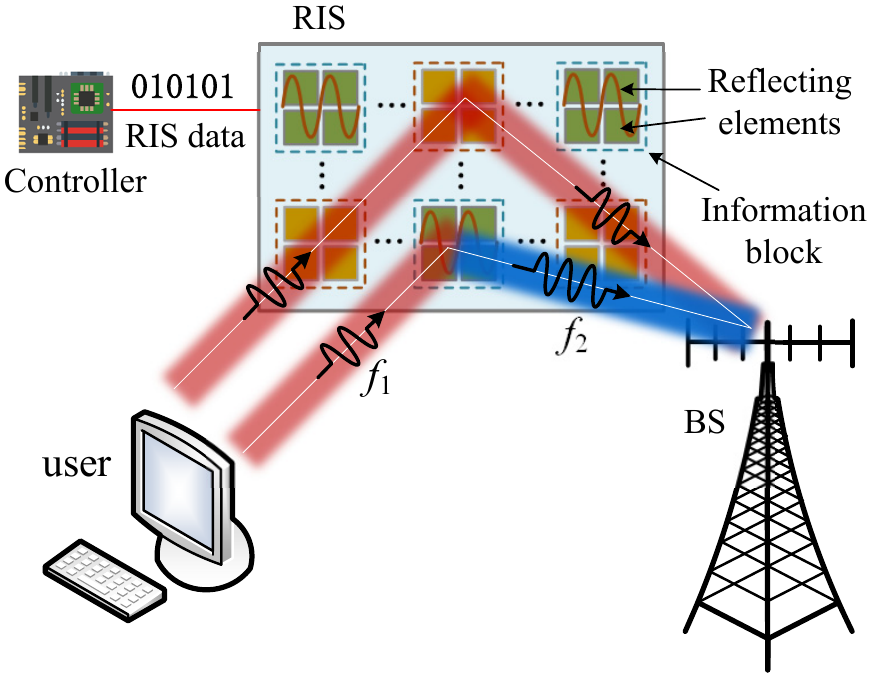}
    \caption{A RIS-aided FRM-OFDM system.}\label{scene}
\end{figure}

As illustrated in Fig.~\ref{scene}, we consider a RIS-aided uplink broadband system employing OFDM, where a single-antenna user communicates with an $M$-antenna BS.
The RIS is equipped with a controller to dynamically manipulate the phase shifts of its $N$ reflection elements. Let $\tilde{\bsm{\theta}}\triangleq [\tilde{\theta}_1,\cdots, \tilde{\theta}_N] $ denote the phase shift vector of the RIS with $|\tilde{\theta}_n|=1, \forall n$. 
Denote by $K$ the number of orthogonal SCs with $K$ being an even integer. The frequency interval between any two adjacent SCs is $\frac{1}{T}$, where $T$ is the time duration of each OFDM symbol without considering the cyclic prefix (CP). We assume that all the user-BS, user-RIS, and RIS-BS links are quasi-static block fading, i.e., the channel remains constant within the channel coherence time (much longer than $T$). Due to multiple scattering with different delays, all three channel links are modeled as frequency-selective fading. 

Assuming that the user-BS, user-RIS, and RIS-BS links have $L_d$, $L_1$ and $L_2$ taps in impulse response, respectively.
The baseband equivalent channel impulse response (CIR) from the user to the $m$th BS antenna (denoted by BS$_m$) is given by
\begin{align}
    g_m^{\rm UB} (t)= \sum_{l=1}^{L_d} g^{\rm UB}_{m,l} \delta(t-\tau^{\rm UB}_l )
\end{align}
where $g^{\rm UB}_{m,l}$ and $\tau^{\rm UB}_l$ are respectively the channel coefficient and the delay of the $l$th tap $\forall l \in \mathcal{I}_{L_d}$. Similarly, the CIRs from the user to the $n$th RIS element (denoted by RIS$_n$), and from RIS$_n$ to BS$_m$ are respectively given by 
\begin{align}
    g_{n}^{\rm UR}(t)= \sum_{i=1}^{L_1} g_{n,i}^{\rm UR} \delta(t-\tau_{i}^{\rm UR} )
\quad \text{and} \quad
g_{m,n}^{\rm RB}(t)= \sum_{j=1}^{L_2} g_{m,n,j}^{\rm RB} \delta(t-\tau_{j}^{\rm RB} ), \forall n \in \mathcal{I}_N,
\end{align}
with $\{ g_{n,i}^{\rm UR}\}$ and  $\{ g_{m,n,j}^{\rm RB}\}$ being the corresponding channel coefficients, and $\{ \tau_{i}^{\rm UR}\}$ and  $\{\tau_{j}^{\rm RB}\}$ being the delays. Following\cite{wu2019intelligent},\cite{huang2019reconfigurable},\cite{li2020reconfigurable},\cite{yang2021ris}, we assume that the channel state information (CSI) is perfectly known by the BS in this paper. The CSI acquisition in RIS-aided OFDM systems can be done, e.g., by employing the recently developed channel estimation techniques in \cite{zheng2019intelligent} and \cite{zheng2020intelligent}.

Each OFDM symbol is represented by $\bsm{x}=[x_1, \cdots , x_{K}]^{\rm T} $, where the user symbol sent over the $k$th SC, $x_k$, is independently and uniformly drawn from a constellation set $\mathcal{X}$, i.e., $p_{\bsm{x}}(\bsm{x}) = \prod_{k=1}^K p_{x_k}(x_k) = \frac{1}{|\mathcal{X}^K|}$.
We assume equal power allocation in different SCs with power constraint $\frac{1}{K} \mbs{E}\bsm{x}\bsm{x}^{\rm H} = P$, where $P$ is the power budget. 
In the transmission of an OFDM symbol, the $K$ signals $\{x_k\}$ are firstly modulated on $K$ orthogonal SCs.  
The generated time-domain OFDM signal after the modulation is $x(t)=  \sum_{k=1}^{K} x_k e^{j2\pi\frac{k-1}{T}t}, t\in [0,T]$.
Then, $x(t)$ is usually appended by a CP of length $L_{cp}$ to suppress the inter-subcarrier interference (ISI), where $L_{cp} \geq \max \left( \{\tau^{\rm UB}_l\}, \{\tau_{i}^{\rm UR}\}\! +\! \{\tau_{j}^{\rm RB} \} \right)$. At the BS side, the received signal at BS$_m$ after removing the CP is
\begin{align} \label{SysMod_1}
    y_m(t)&= g_m^{\rm UB}(t) \circledast x(t) + \sum_{n=1}^N g_{m,n}^{\rm RB}(t) \circledast  \left(\tilde{\theta}_n(t) \left( g_{n}^{\rm UR}(t) \circledast x(t) \right)\right)  + w(t),  t\in [0,T],
\end{align} 
where $w(t)$ is an additive white Gaussian noise.\footnote{Note that if the bandwidth of the OFDM system is large, the phase and amplitude responses of a RIS may vary within the bandwidth. In this paper, for ease of exposition, we consider an OFDM system with an appropriate bandwidth within which the phase and amplitude responses of the RIS can be regarded as constants. We leave the study of varying responses of phase and amplitude on the RIS at different SCs to future work.}

\subsection{Frequency Reflection Modulation}  \label{Sec.IIBB}

In most existing RIS-aided systems, the phases of all RIS elements remain constant within the duration of at least one time slot.
The FRM-OFDM scheme, by contrast, exploits the feature that the RIS phase shifts can be tuned continuously over time to achieve the PIT of RIS. 
Specifically, in the FRM-OFDM scheme, the phase of the $n$th RIS element $\forall n \in \mathcal{I}_N$ is given by
\begin{align} \label{theta_t}
    \tilde{\theta}_n(t) = \begin{cases}
        \theta_n & {\rm if}~ s_n=1 \\
        \theta_n e^{j\left(2\pi\frac{t}{T}\right)} & {\rm if}~ s_n=0
      \end{cases}
\end{align}
where the auxiliary variable $s_n \in [0, 1], \forall n \in \mathcal{I}_n$ is defined to indicate the phase state of the $n$th element: $s_n=0$ means the ``frequency-hopping'' state and $s_n=1$ means the ``no-hopping'' state. In the no-hopping state, $\tilde{\theta}_n(t), t \in [0,T]$ is set to the optimized phase shift $\theta_n\triangleq e^{j \psi_n}, \psi_n \in [0,2\pi), \forall n \in \mathcal{I}_n$ under a certain passive beamforming design. In the frequency-hopping state, apart from the system optimization component $\theta_n$, $\tilde{\theta}_n(t)$ has another component $e^{j2\pi\frac{t}{T}}$ that modulates the impinging EM waves with frequency $\frac{1}{T}$. Recall that $\frac{1}{T}$ is exactly the frequency interval between any two adjacent SCs. The phase of each RIS element is switched between the two states according to the RIS date. Thus, the receiver can retrieve the RIS information by detecting its phase states.

Based on the above configuration of the RIS, we next give mathematical models to characterize the RIS-aided FRM-OFDM system. We start with the channel response of the RIS-aided OFDM system in a noiseless case with $s_n=0, \forall n \in \mathcal{I}_n$ and $g_m^{\rm UB}(t)=0$ (i.e., the direct link is blocked by obstacles). In this case, the received signal at the BS$_m$ is 
\begin{align} \label{SysMod_2}
    y_m(t)&= \sum_{n=1}^N g_{m,n}^{\rm RB}(t) \circledast \left(\theta_ne^{j2\pi\frac{t}{T}}  \left( g_{n}^{\rm UR}(t) \circledast x(t) \right) \right), t \in [0,T].
\end{align}
The demodulated OFDM output at the $k$th SC of BS$_m$ is given by
\begin{align}  \label{SysMod_3}
    y_{k,m}&= \int_{0}^{T} y_m(t) e^{-j\frac{2\pi}{T}{(k-1)}t} \,dt  \notag\\
    &= \sum_{n=1}^N \theta_n \int_{0}^{T} \int_{0}^{T} g_{m,n}^{\rm RB}(\tau) \left(e^{j\frac{2\pi}{T}(t-\tau)}\left( g_{n}^{\rm UR}(t-\tau)\circledast x(t-\tau)\right) \right) e^{-j\frac{2\pi}{T}(k-1)t} \,d\tau \,dt  \notag\\
    &= \sum_{n=1}^N \theta_n \int_{0}^{T} g_{m,n}^{\rm RB}(\tau) e^{-j\frac{2\pi}{T}(k-1) \tau} \,d\tau \int_{0}^{T} e^{j\frac{2\pi}{T}(t-\tau)} \left( g_{n}^{\rm UR}(t-\tau)\circledast x(t-\tau)\right) e^{-j\frac{2\pi}{T}(k-1)(t-\tau)} \,d(t-\tau)  \notag\\
    &= \sum_{n=1}^N \theta_n h_{k,m,n}^{\rm RB} \int_{0}^{T}\left( g_{n}^{\rm UR}(t)\circledast x(t)\right) e^{-j\frac{2\pi}{T}(k-2)t} \,dt = \sum_{n=1}^N \theta_n h_{k,m,n}^{\rm RB}  h_{k-1,n}^{\rm UR} x_{k-1}, 
\end{align}
where $h_{k,m,n}^{\rm RB}= \int_{0}^{T} g_{m,n}^{\rm RB}(\tau) e^{-j\frac{2\pi}{T}(k-1) \tau} \,d\tau $ is the frequency response of $g_{m,n}^{\rm RB}(t)$ at the $k$th SC, and $h_{k-1,n}^{\rm UR} = \int_{0}^{T} g_{n}^{\rm UR}(\tau) e^{-j\frac{2\pi}{T}(k-2)\tau}\,d\tau$ is the frequency response of $g_{n}^{\rm UR}(t)$ at the $(k-1)$th SC, $\forall n $, $\forall k$.  Eq.~\eqref{SysMod_3} shows that, under the setting of $s_n=0, \forall n \in \mathcal{I}_n$, the received signal of BS$_m$ at the $k$th SC (i.e., $y_{k,m}$) is the observation of the signal sent by the $(k-1)$th SC  (i.e., $x_{k-1}$).  This can be explained by the modulation effect of $e^{j2\pi\frac{1}{T}t}$.

In the FRM-OFDM system, since each RIS element is randomly taken in a frequency-hopping or no-hopping state, the demodulated receive signal of BS$_m$ at the $k$th SC is represented by 
\begin{align}  \label{SysMod_4} 
    y_{k,m} = &\sum_{n=1}^N\left(x_k h_{k,m,n}^{\rm RB} h_{k,n}^{\rm UR}\theta_ns_n \!+\! x_{k-1} h_{k,m,n}^{\rm RB} h_{k-1,n}^{\rm UR}\theta_n (1-s_n) \right)   \notag\\
          &+ x_k h_{k,m}^{\rm UB}    + w_{k,m}, \quad \forall (k - 1)\in \mathcal{I}_{K-1},    
\end{align}
where $h_{k,m}^{\rm UB} = \int_{0}^{T} g_m^{\rm UB}(\tau) e^{-j\frac{2\pi}{T}(k-1) \tau} \,d\tau $ is the channel coefficient of the direct link, and the Gaussian noise $w_{k,m}$ obeys $\mathcal{CN}(;0,\sigma^2), \forall k \in \mathcal{I}_{k}$.

In the FRM-OFDM system, the modulation operation at the RIS can shift the signal modulated on the $K$th SC to an out-of-band frequency, which leads to spectrum leakage. To avoid this phenomenon, we only transmit signals on the first $K-1$ SCs and leave the $K$th SC blank, i.e.,  $\bsm{x}=[x_1, \cdots , x_{K-1}]^{\rm T} $.
For notational convenience, we introduce auxiliary variables $x_0 = 0$ and $x_{K} = 0$.   
Denote by $\bsm{h}_{k}^{\rm UB} \triangleq [h_{k,1}^{\rm UB}, \cdots, h_{k,M}^{\rm UB} ]^{\rm T} \in\mathbb{C}^{M}$, $\bsm{h}_k^{\rm UR} \triangleq  [h_{k,1}^{\rm UR}, \cdots, h_{k,N}^{\rm UR}]^{\rm T} \in\mathbb{C}^{N}$, and $\bsm{h}_{k,m}^{\rm RB} \triangleq  [h_{k,m,1}^{\rm RB}, \cdots, h_{k,m,N}^{\rm RB}]^{\rm T} \in\mathbb{C}^{N}$ the frequency-domain channels of the user-BS link, the user-RIS link, and the RIS-BS$_m$ link at the $k$th SC, respectively. $\bsm{H}_k^{\rm RB} \triangleq [\bsm{h}_{k,1}^{\rm RB}, \cdots, \bsm{h}_{k,M}^{\rm RB}]^{\rm T} \in\mathbb{C}^{M \times N} $. Define $\bsm{w}_k = [w_{k,1}, \cdots, w_{k,M}]^{\rm T}$. Then, the received signal at the $k$th SC is
\begin{align}  \label{SysMod_5} 
    \bsm{y}_{k} &= x_k (\bsm{h}_{k}^{\rm UB} + \underbrace{ \bsm{H}_k^{\rm RB} \diag\{\bsm{h}_k^{\rm UR}\}}_{\triangleq \bsm{H}_k} \bsm{\Theta}\bsm{s}) + x_{k-1} \underbrace{\bsm{H}_k^{\rm RB}\diag\{\bsm{h}_{k-1}^{\rm UR}\}}_{\triangleq \tilde{\bsm{H}}_{k}} \bsm{\Theta}  (\textbf{1}_{N}-\bsm{s}) + \bsm{w}_k, \forall k,
\end{align}
where  $\bsm{\Theta} = \diag\{\bsm{\theta}\}$, and $\bsm{s} = [ s_1, \cdots, s_N]^{\rm T}$. Define $\bsm{y} \triangleq [\bsm{y}_{1}^{\rm T}, \cdots, \bsm{y}_{K}^{\rm T}]^{\rm T } $, $\bsm{h}^{\rm UB} \triangleq [(\bsm{h}_1^{\rm UB})^{\rm T}, \cdots, (\bsm{h}_K^{\rm UB})^{\rm T}]^{\rm T}$,
$\bsm{H} \triangleq [\bsm{H}_{1}^{\rm T}, \cdots, \bsm{H}_{K-1}^{\rm T}, \mathbf{0}_{N \times M}]^{\rm T } \in\mathbb{C}^{MK \times N}$, $\tilde{\bsm{H}} \triangleq [\mathbf{0}_{N \times M},\tilde{\bsm{H}}_{2}^{\rm T}, \cdots, \tilde{\bsm{H}}_{K}^{\rm T}]^{\rm T } \in\mathbb{C}^{MK \times N}$, and $\bsm{w} \triangleq [\bsm{w}_{1}^{\rm T}, \cdots, \bsm{w}_{K}^{\rm T}]^{\rm T } \in\mathbb{C}^{MK}$. Then, the FRM-OFDM system can be modelled as 
\begin{align}  \label{SysMod_6} 
    \bsm{y} &= (\diag\{[\bsm{x}^{\rm T},0]\} \otimes \mathbf{I}_M) (\bsm{h}^{\rm UB} + \bsm{H}\bsm{\Theta}\bsm{s}) + (\diag\{[0,\bsm{x}^{\rm T}]\} \otimes \mathbf{I}_M) \tilde{\bsm{H}} \bsm{\Theta}(\textbf{1}_{N}-\bsm{s}) + \bsm{w}. 
\end{align}
 We further adopt an element-grouping strategy \cite {yang2020intelligent, zheng2019intelligent,zheng2020intelligent} to achieve the flexible transmission rate for the RIS.
To be specific, the $N$ elements of the RIS are divided into $B$ blocks, $B\in \mathcal{I}_N$. Each of them consists of adjacent $L = \lfloor \frac{N}{B}\rfloor $\footnote{If $\frac{N}{B}$ is not an integer, the residual ($N$ mod $B$) elements of the RIS are assigned to the last block. For notational convenience, we assume  ($N$ mod $B=0$) in the following.} elements sharing a common phase state, denoted by $\bsm{c} = [c_1,\cdots,c_B]^{\rm T}$ where $\{c_b\}$ are independently and uniformly drawn from $\{0,1\}$. Thus, the information carried by each block $c_b $ is one bit. By changing $B$, the transmission rate of the RIS ranges from $0$ to $N$ bits. 
Then, we have $\bsm{s} = \bsm{c} \otimes \textbf{1}_L = \bsm{T}\bsm{c}$,
where $\bsm{T} \triangleq \textbf{I}_B \otimes \textbf{1}_L $.

\subsection{Advantages and Challenges} \label{Sec.IIC}

The RIS-aided FRM-OFDM scheme is advantageous over the existing ORM schemes in many aspects. As mentioned in the Introduction, the ORM schemes encode the RIS data into the on/off states of the RIS elements, which compromises the PB capability of RIS. The model of the ORM-OFDM scheme can be readily obtained by omitting the second component of \eqref{SysMod_6}.
Clearly, the FRM-OFDM scheme retains the signals reflected by the ``turned-off'' elements (those corresponding to $s_n = 0, \forall n$). 
Another advantage of the FRM-OFDM scheme is in the outage probability. The ORM scheme proposed in \cite{yan2019passive} has a link-outage risk because the number of the ``turned-on'' elements cannot be guaranteed. This also leads to the fluctuation of the reflected signal power. By contrast, in FRM-OFDM, the BS can receive the signals reflected by all the RIS elements, thereby avoiding these two problems. 

However, the FRM-OFDM scheme faces two main challenges in system design. First, due to the simultaneous transmission of the user and RIS information, the FRM-OFDM system is an MMAC. Characterizing the capacity of the MMAC is a challenging task. Second, the joint recovery of the user symbols and the RIS data at the BS gives rise to a bilinear detection problem. In the subsequent two sections, 
we take a first attempt to address the above challenges.
\section{RIS Phase Shift Design} \label{Sec.III}
\subsection{Problem Formulation} \label{Sec.III.I}

From information theory, the capacity of the FRM-OFDM system is the closure of the convex hull of all $(R_{\bsm{x}}, R_{\bsm{s}})$, satisfying \cite{cover2012elements}

\begin{subequations}
    \begin{align}
        R_{\bsm{x}} &\leq I(\bsm{x};\bsm{y}|\bsm{s}); \\
        R_{\bsm{s}} &\leq I(\bsm{s};\bsm{y}|\bsm{x}); \\
        R_{\bsm{x}} + R_{\bsm{s}} &\leq I(\bsm{x},\bsm{s};\bsm{y}),
    \end{align}
\end{subequations} 
where $R_{\bsm{x}}$ and $R_{\bsm{s}}$ are respectively the information rates of the user and the RIS;
 $I(\bsm{x};\bsm{y}|\bsm{s})$ is the conditional mutual information between $\bsm{x}$ and $\bsm{y}$ conditioned on $\bsm{s}$, and vice verse; and $I(\bsm{x},\bsm{s};\bsm{y})$ is the mutual information between $(\bsm{x}, \bsm{s})$ and $\bsm{y}$. In this paper, we aim to maximize the sum rate of user and RIS over the RIS phase shift $\bsm{\theta}$, i.e.,
\begin{align}
\mathcal{P}_1 \quad \max_{\bsm{\theta}} \quad& I(\bsm{x},\bsm{s};\bsm{y}) \notag \\ 
\textrm{s.t.}\quad&|\theta_n| = 1, \forall n. \label{opt.obj} 
\end{align} 
However, $I(\bsm{x},\bsm{s};\bsm{y})$ is difficult to compute because it involves a multi-dimensional integration over the probability density functions of $\bsm{x}$ and $\bsm{s}$. Previous papers \cite{yan2019passive,yan2020passive,lin2020reconfigurable}  simplify this problem by approximating $I(\bsm{x},\bsm{s};\bsm{y}) = I(\bsm{x};\bsm{y}|\bsm{s}) + I(\bsm{s};\bsm{y})$ as $I(\bsm{x};\bsm{y}|\bsm{s})$, which is a lower bound of $I(\bsm{x},\bsm{s};\bsm{y})$ by noting $I(\bsm{s};\bsm{y}) \geq 0$. To obtain more insights into $I(\bsm{x},\bsm{s};\bsm{y})$, it is important to take account $I(\bsm{s};\bsm{y})$ in RIS phase shift optimization. Thus, in this paper, we directly cope with the maximization of the sum rate of the FRM-OFDM system.

We first decompose $I(\bsm{x},\bsm{s};\bsm{y})$ as $I(\bsm{x},\bsm{s};\bsm{y}) = H(\bsm{y}) - H(\bsm{y}|\bsm{x},\bsm{s}) = H(\bsm{y}) - H(\bsm{w})$, where $H(\bsm{y})$ and $H(\bsm{y}|\bsm{x},\bsm{s})$ are respectively the entropy and the conditional entropy of $\bsm{y}$, and $H(\bsm{w})$ is the entropy of the noise. Thus, maximizing $I(\bsm{x},\bsm{s};\bsm{y})$ over $\bsm{\theta}$ is equivalent to
$ \max_{\bsm{\theta}} H(\bsm{y}) = \max_{\bsm{\theta}} -p_{\bsm{y}}(\bsm{y}) \log(p_{\bsm{y}}(\bsm{y}))$.
However, the distribution of $\bsm{y}$ is very complicated by recalling the sophisticated channel model given in \eqref{SysMod_6}. Note that $\bsm{y}$ is a mixture of many independent components of $\bsm{x}$ and $\bsm{s}$. Thus, we approximate $p_{\bsm{y}}(\bsm{y})$ as a Gaussian distribution based on the central limit theorem (CLT). This Gaussian approximation generally provides an upper bound to the true entropy of $p_{\bsm{y}}(\bsm{y})$. It is known that this bound is tight in the low-SNR and large-constellation-set regimes \cite{verdu2002spectral,rao2004analysis}. 

With the Gaussian approximation, we have $H(\bsm{y}) = \log(\pi^{MK}e) + \log\det(\bsm{\mathcal{Q}}) $, where 
\begin{align}   \label{opt.Q}
    \bsm{\mathcal{Q}} &=\mbs{E}(\bsm{y} \bsm{y}^{\rm H}) \notag\\
    &= P\left[ \bsm{h}^{\rm UB}  (\bsm{h}^{\rm UB} )^{\rm H} + \bsm{H} \bsm{\Theta} \bsm{\mathcal{C}}_{\bsm{s}\bsm{s}} \bsm{\Theta}^{\rm H} \bsm{H}^{\rm H} + \bsm{h}^{\rm UB}  (\bsm{H} \bsm{\Theta}\mbs{E}\bsm{s})^{\rm H} + \bsm{H} \bsm{\Theta}(\mbs{E}\bsm{s})(\bsm{h}^{\rm UB} )^{\rm H} \right]_{\rm diag}^{M\times M}  \notag\\
    & + P\left[ \tilde{\bsm{H}} \bsm{\theta}\bsm{\theta}^{\rm H} \tilde{\bsm{H}}^{\rm H}  + \tilde{\bsm{H}} \bsm{\Theta} \bsm{\mathcal{C}}_{\bsm{s}\bsm{s}} \bsm{\Theta}^{\rm H} \tilde{\bsm{H}}^{\rm H} - \tilde{\bsm{H}}\bsm{\theta} (\tilde{\bsm{H}}\bsm{\Theta}\mbs{E}\bsm{s})^{\rm H}- \tilde{\bsm{H}} \bsm{\Theta} (\mbs{E}\bsm{s})(\tilde{\bsm{H}}\bsm{\theta})^{\rm H} \right]_{\rm diag}^{M\times M}  \notag\\
    & + P\left[\bsm{h}^{\rm UB}  (\tilde{\bsm{H}} \bsm{\Theta}(\textbf{1}_{N}-\mbs{E}\bsm{s}))^{\rm H}  + \bsm{H} \bsm{\Theta} (\mbs{E}\bsm{s}) (\tilde{\bsm{H}} \bsm{\theta})^{\rm H} - \bsm{H} \bsm{\Theta} \bsm{\mathcal{C}}_{\bsm{s}\bsm{s}} \bsm{\Theta}^{\rm H} \tilde{\bsm{H}}^{\rm H}\right]_{\rm diag+}^{M\times M}  \notag\\
    &  + P\left[\tilde{\bsm{H}} \bsm{\Theta}(\textbf{1}_{N}-\mbs{E}\bsm{s}) (\bsm{h}^{\rm UB} )^{\rm H} + \tilde{\bsm{H}} \bsm{\theta}(\bsm{H} \bsm{\Theta} \mbs{E}\bsm{s})^{\rm H} - \tilde{\bsm{H}} \bsm{\Theta} \bsm{\mathcal{C}}_{\bsm{s}\bsm{s}} \bsm{\Theta}^{\rm H} \bsm{H}^{\rm H} \right]_{\rm diag-}^{M\times M} + \sigma^2 \mathbf{I}_{MK}, 
\end{align}
with $\bsm{\mathcal{C}}_{\bsm{s}\bsm{s}} = \mbs{E} (\bsm{s} \bsm{s}^{\rm T} ) = \frac{1}{4} \textbf{1}_{ N\times N} + \frac{1}{4} \textbf{I}_{N} $, and $\mbs{E} \bsm{s} = \frac{1}{2}\textbf{1}_{N} $. It is worth noting that $\bsm{\mathcal{Q}}$ is a block tri-diagonal mtrix with block size $M \times M$.
Define a block bi-diagonal matrix $\bsm{H}^{\bsm{\theta}} \in \mathbb{C}^{MK\times (K-1)}$ with the $k$th blocks on the diagonal and the lower sub-diagonal being $\bsm{h}_k^{\rm UB}  + \frac{1}{2}\bsm{H}_k \bsm{\theta} \in \mathbb{C}^M$ and $\frac{1}{2}\tilde{\bsm{H}}_{k+1} \bsm{\theta}\in \mathbb{C}^M$, respectively. Then, $\bsm{\mathcal{Q}}$ can be further expressed as 
\begin{align}  
    \bsm{\mathcal{Q}} & =  \bsm{\mathcal{Q}}^{\bsm{\theta}} +  \bsm{\mathcal{Q}}^{/ \bsm{\theta}},  \label{Qy}
\end{align}
where $ \bsm{\mathcal{Q}}^{\bsm{\theta}} \triangleq P\bsm{H}^{\bsm{\theta}} (\bsm{H}^{\bsm{\theta}})^{\rm H} $ and $\bsm{\mathcal{Q}}^{/ \bsm{\theta}} \triangleq \frac{P}{4} \left[\bsm{H} \bsm{H}^{\rm H} + \tilde{\bsm{H}} \tilde{\bsm{H}}^{\rm H}\right]_{\rm diag}^{M\times M} - \frac{P}{4}\left[\bsm{H}\tilde{\bsm{H}}^{\rm H}\right]_{\rm diag+}^{M\times M} - \frac{P}{4}\left[\tilde{\bsm{H}}\bsm{H}^{\rm H}\right]_{\rm diag-}^{M\times M} + \sigma^2 \mathbf{I}_{MK}$.

With the above results, we recast $\mathcal{P}_1$ as 
\begin{align}
    \mathcal{P}_2 \quad \max_{\bsm{\theta}} \quad& \log\det(\bsm{\mathcal{Q}}^{\bsm{\theta}} + \bsm{\mathcal{Q}}^{\bsm{/ \theta}} ) \notag \\ 
    \textrm{s.t.}\quad& \eqref{opt.obj}.
\end{align}
$\mathcal{P}_2$ can be approximately solved by the semi-definite program (SDP) by converting it to a homogeneous quadratically constrained quadratic program (QCQP) \cite{luo2010sdp}. However, the computational complexity of the SDP is $O(N^6)$, which is high as the number of RIS elements $N$ is large in practice. We next propose an alternating optimization (AO) algorithm to efficiently solve $\mathcal{P}_2$.  

\subsection{Alternating Optimization Algorithm} 

We design our AO algorithm by first introducing an equivalent RIS-aided MIMO channel for the MMAC in the FRM-OFDM system in the sense of RIS phase shift optimization.
Note that $\bsm{\mathcal{Q}}$ in \eqref{Qy} can be regarded as the covariance matrix of the auxiliary system: 
\begin{align} \label{Equ.Sys}
    \bsm{y} & = \bsm{H}^{\bsm{\theta}} \bsm{x} + \bsm{w}^{\rm equ},
\end{align}
where $\bsm{w}^{\rm equ}$ is an independent colored noise with covariance matrix $\bsm{\mathcal{Q}}^{/ \bsm{\theta}} $. Thus, solving $\mathcal{P}_2$ is equivalent to maximizing the rate of the system in \eqref{Equ.Sys}, i.e., $\max_{\bsm{\theta}} I(\bsm{x}; \bsm{y})$. This equivalent replacement has great significance in reducing the complexity of the RIS phase shift design in FRM-OFDM systems, because the  rate maximization in a MIMO channel is more tractable, and there exist effective existing algorithms, such as the the WMMSE algorithm \cite{shi2011iteratively}, to solve the problem.

The WMMSE algorithm reformulates the sum rate maximization in $\mathcal{P}_2$ as
\begin{align} 
    \mathcal{P}_3 \quad &\max_{\bsm{\theta}, \bsm{\Sigma}, \bsm{\Phi}} f(\bsm{\theta}, \bsm{\Sigma}, \bsm{\Phi}) \notag \\ 
    \textrm{s.t.} \quad&\eqref{opt.obj}
\end{align}
where $f(\bsm{\theta}, \bsm{\Sigma}, \bsm{\Phi}) \triangleq \log\det(\bsm{\Sigma}) - \left\|\bsm{\Sigma}^{\frac{1}{2}} (\bsm{x} -\bsm{\Phi}\bsm{y})\right\|^2$, and $\bsm{\Sigma} \succeq \mathbf{0}$ and $\bsm{\Phi}$ are auxiliary variables. $\mathcal{P}_3$ is a non-convex problem. But for a given $\bsm{\theta}$, $f(\bsm{\Sigma}, \bsm{\Phi})$ is concave in $(\bsm{\Sigma}, \bsm{\Phi})$. Thus, we next optimize $\bsm{\theta}$ and $(\bsm{\Sigma}, \bsm{\Phi})$ in an alternating way. 

\subsubsection{Optimizing $(\bsm{\Sigma}, \bsm{\Phi})$ for given $\bsm{\theta}$} 
For given $\bsm{\theta}$, $\mathcal{P}_3$ is reduced to 
\begin{align}
    \mathcal{P}_{3.1} \quad \max_{\bsm{\Sigma}, \bsm{\Phi}} f(\bsm{\Sigma}, \bsm{\Phi}).
\end{align}
$\mathcal{P}_{3.1}$ is a concex problem and the optimal $\bsm{\Sigma}$ and $\bsm{\Phi}$ can be obtained by taking the first-order derivative, given as
\begin{align} 
    \bsm{\Sigma} = \left( \bsm{\mathcal{C}}_{\bsm{x}\bsm{x}} - \bsm{\mathcal{C}}_{\bsm{x}\bsm{y}} \bsm{\mathcal{C}}_{\bsm{y}\bsm{y}}^{-1} \bsm{\mathcal{C}}_{\bsm{y}\bsm{x}} \right)^{-1} \quad \text{and} \quad\bsm{\Phi} = \bsm{\mathcal{C}}_{\bsm{x}\bsm{y}} \bsm{\mathcal{C}}_{\bsm{y}\bsm{y}}^{-1},  \label{opt.phi}
\end{align}
where $\bsm{\mathcal{C}}_{\bsm{x}\bsm{x}} = \mbs{E} (\bsm{x}\bsm{x}^{\rm H}) = P\mathbf{I}_{K-1} $, $\bsm{\mathcal{C}}_{\bsm{x}\bsm{y}} = \mbs{E} \left(\bsm{x}\bsm{y}^{\rm H}\right) = \bsm{\mathcal{C}}_{\bsm{x}\bsm{x}} (\bsm{H}^{\bsm{\theta}})^{\rm H} $, $\bsm{\mathcal{C}}_{\bsm{x}\bsm{y}} = \bsm{\mathcal{C}}_{\bsm{y}\bsm{x}}^{\rm H}$, and $\bsm{\mathcal{C}}_{\bsm{y}\bsm{y}} = \mbs{E} \left(\bsm{y}\bsm{y}^{\rm H}\right) = \bsm{\mathcal{Q}} $. 

\subsubsection{Optimizing $\bsm{\theta}$ for given $(\bsm{\Sigma}, \bsm{\Phi})$}
For given $(\bsm{\Sigma}, \bsm{\Phi})$, the maximization of $f(\bsm{\theta})$ is to solve 
\begin{align}
    & \min_{\bsm{\theta}} \|\bsm{\Sigma}^{\frac{1}{2}} (\bsm{x} -\bsm{\Phi}\bsm{y} \|^2 = \min_{\bsm{\theta}} \tr \left\{ (\bsm{H}^{\bsm{\theta}})^{\rm H} \bsm{\Phi}^{\rm H} \bsm{\Sigma} \bsm{\Phi}\bsm{H}^{\bsm{\theta}}- 2\Re\{\bsm{\Sigma} \bsm{\Phi}\bsm{H}^{\bsm{\theta}}\} \right\}. \label{Opt.f}
\end{align}
The first term in \eqref{Opt.f} can be derived as 
\begin{align}
\min_{\bsm{\theta}} \bsm{\theta}^{\rm H} \bsm{\Lambda} \bsm{\theta}  + 4\Re\left\{ \left(\bsm{h}^{\rm UB}\right)^{\rm H} \left([\bsm{\Phi}^{\rm H} \bsm{\Sigma} \bsm{\Phi}]^{M \times M}_{\rm diag} \bsm{H} + [\bsm{\Phi}^{\rm H} \bsm{\Sigma} \bsm{\Phi}]^{M \times M}_{\rm diag+}\tilde{\bsm{H}}\right)\bsm{\theta} \right\} 
\end{align}
where 
\begin{align}
    \bsm{\Lambda} &\triangleq \bsm{H}^{\rm H} [\bsm{\Phi}^{\rm H} \bsm{\Sigma} \bsm{\Phi}]^{M \times M}_{\rm diag} \bsm{H} + \tilde{\bsm{H}}^{\rm H} [\bsm{\Phi}^{\rm H} \bsm{\Sigma} \bsm{\Phi}]^{M \times M}_{\rm diag} \tilde{\bsm{H}}  \notag\\
    &+ \bsm{H}^{\rm H} [\bsm{\Phi}^{\rm H} \bsm{\Sigma} \bsm{\Phi}]^{M \times M}_{\rm diag+}\tilde{\bsm{H}} + \tilde{\bsm{H}}^{\rm H} [\bsm{\Phi}^{\rm H} \bsm{\Sigma} \bsm{\Phi}]^{M \times M}_{\rm diag-}\bsm{H}. \label{opt.Lambda} 
\end{align}
The second term in \eqref{Opt.f} can be derived as
\begin{align}
    \min_{\bsm{\theta}} 2\Re\left\{\mathbf{1}_{K-1}^{\rm T} \left([\bsm{\Sigma} \bsm{\Phi}]^{1 \times M}_{\rm diag}\bsm{H} + [\bsm{\Sigma} \bsm{\Phi}]^{1 \times M}_{\rm diag+}\tilde{\bsm{H}} \right)\bsm{\theta} \right\}.
\end{align}
Denote 
\begin{align}
    \bsm{\alpha} &\triangleq 2 \left(\bsm{h}^{\rm UB}\right)^{\rm H} \left([\bsm{\Phi}^{\rm H} \bsm{\Sigma} \bsm{\Phi}]^{M \times M}_{\rm diag} \bsm{H} + [\bsm{\Phi}^{\rm H} \bsm{\Sigma} \bsm{\Phi}]^{M \times M}_{\rm diag+}\tilde{\bsm{H}}\right) \notag\\
    &- \mathbf{1}_{K-1}^{\rm T} \left([\bsm{\Sigma} \bsm{\Phi}]^{1 \times M}_{\rm diag}\bsm{H} + [\bsm{\Sigma} \bsm{\Phi}]^{1 \times M}_{\rm diag+}\tilde{\bsm{H}}\right).  \label{opt.alpha} 
\end{align} 
Then, the objective of optimizing $\bsm{\theta}$ for given $(\bsm{\Sigma}, \bsm{\Phi})$ is converted to 
\begin{align}
    \mathcal{P}_{3.2} \quad \min_{\bsm{\theta}} \quad& \bsm{\theta}^{\rm H} \bsm{\Lambda}  \bsm{\theta} 
    + 2\Re\left\{\bsm{\theta}^{\rm H} \bsm{\alpha} \right\}  \notag \\ 
    \textrm{s.t.}\quad& \eqref{opt.obj}. 
\end{align}
$\mathcal{P}_{3.2}$ is a QCQP with non-convex unit-modulus constraint \eqref{opt.obj}. This problem has been extensively investigated in the field of RIS phase shift optimization. The existing algorithms to solve this problem include the SDR \cite{wu2019Towards} and majorization-minimization (MM) algorithms \cite{pan2020multicell}. Because the computational complexity of SDP is high, we choose the MM algorithm for solving $\mathcal{P}_{3.2}$ in this paper.   

\begin{algorithm}[t]
    \caption{ The MM Algorithm}
    \label{alg:MM}
    \begin{algorithmic}[1]
    \REQUIRE $\bsm{\Lambda}$, $\bsm{\alpha}$
    \STATE Initialize $\bsm{\theta}$ randomly
    \REPEAT
    \STATE Compute $\bsm{q}= \left(\lambda_{\text{max}} \textbf{I}_{N} - \bsm{\Lambda} \right)\bsm{\theta} - \bsm{\alpha} $
    \STATE Update $\bsm{\theta} = e^{j \arg(\bsm{q})}$
    \UNTIL convergence criterion is met  
    \ENSURE $\bsm{\theta}$
    \end{algorithmic}
 \end{algorithm}  
 The key idea of the MM algorithm is to repeatedly minimize an auxiliary function that upper-bounds the original objective function. 
The update rules of this algorithm are provided in Algorithm~\ref{alg:MM}, where $\lambda_{\text{max}}$ in Line $3$ is the maximum eigenvalue of $\bsm{\Lambda}$, $\bsm{q}$ is an intermediate variable, and $\arg(\cdot)$ in Line $4$ returns the argument(s) of the input in an element-wise manner. For details of the MM algorithm please refer to \cite{pan2020multicell}.  

\begin{algorithm}[t]
    \caption{The AO Algorithm }
    \label{alg_AO}
\begin{algorithmic}[1]
   \REQUIRE $\bsm{h}^{\rm UB}$, $\bsm{H}$, $\tilde{\bsm{H}}$, and $\bsm{\mathcal{Q}}$
    \STATE Initialize $\bsm{\theta}$ randomly
    \REPEAT
    \STATE Update $(\bsm{\Sigma}, \bsm{\Phi})$ according to \eqref{opt.phi}
    \STATE Update $(\bsm{\Lambda}, \bsm{\alpha})$ according to \eqref{opt.Lambda} and \eqref{opt.alpha}, respectively
    \STATE Update $\bsm{\theta}$ by the MM algorithm given in Algorithm \ref{alg:MM}
    \UNTIL convergence criterion is met 
    \ENSURE $\bsm{\theta}$ 
\end{algorithmic}  
\end{algorithm}

The overall AO algorithm is summarized in Algorithm \ref{alg_AO}. 
As the target function of $\mathcal{P}_3$ decreases monotonically in each update,  
the convergence of Algorithm~\ref{alg_AO} is guaranteed. 
We now analyze the computational complexity of the proposed AO algorithm. The complexity in Step $3$ is dominated by the matrix inversion of $\bsm{\mathcal{C}}_{\bsm{y}\bsm{y}}$ with complexity $\mathcal{O}((KM)^3)$. The complexity in Step $4$ is $\mathcal{O}(NK^2M^2)$. The complexity of the MM algorithm in Step $5$ is $\mathcal{O}(N^3 + \tau N^2)$ \cite{pan2020multicell}, where $\tau$ is the number of iterations of MM.

\subsection{Recursive Alternating Optimization Algorithm}

The AO algorithm is computationally involving due to the matrix inversion of $\bsm{\mathcal{C}}_{\bsm{y}\bsm{y}}$. To further reduce the computational complexity, we propose a recursive AO (RAO) algorithm by decomposing the mutual information between $\bsm{x}$ and $\bsm{y}$ in a recursive manner. First, we approximate the noise covariance matrix $\bsm{\mathcal{Q}}^{/\bsm{\theta}}$ as a block diagonal matrix with block size $M\times M$ by omitting its off-diagonal blocks. Then, the channels of different SCs are irrelevant with each other. As the signal $x_k$ is only transmitted over the $k$th and the $(k+1)$th SCs, $\forall k \in \mathcal{I}_{K-1}$, the observation for each $x_k$ is 
\begin{align}  
    \left[\begin{array}{c}
        \bsm{y}_k \\ \bsm{y}_{k+1}
        \end{array}\right] &= \left[\begin{array}{c}
            \bsm{h}_k^{\rm UB} + \frac{1}{2}\bsm{H}_k\bsm{\theta}  \\ \frac{1}{2}\tilde{\bsm{H}}_{k+1}\bsm{\theta} 
    \end{array}\right] x_k
    + \left[\begin{array}{c}
        \frac{1}{2}\tilde{\bsm{H}}_k\bsm{\theta}x_{k-1} + \bsm{w}_k^{\rm equ}   \\  \tilde{\bsm{w}}_{k+1}^{\rm equ}
        \end{array}\right],
\end{align}
where the covariance matrix of $\bsm{w}_k^{\rm equ}$ is the $k$th diagonal block of $\bsm{\mathcal{Q}}^{/\bsm{\theta}}$ (denoted by $\bsm{\mathcal{Q}}^{/\bsm{\theta}}_k$), and $\tilde{\bsm{w}}_{k+1}^{\rm equ} \triangleq \left(\bsm{h}_{k+1}^{\rm UB} + \frac{1}{2}\bsm{H}_{k+1}\bsm{\theta}\right) x_{k+1} + \bsm{w}_{k+1}^{\rm equ}$. 
Based on the chain rule of mutual information \cite{cover2012elements}, $I(\bsm{x}; \bsm{y}) $ can be decomposed as
\begin{align} 
    &I(\bsm{x}; \bsm{y}) = \sum_{k=1}^{K-1} I(x_k;\bsm{y}|x_1,\cdots,x_{k-1}) = \sum_{k=1}^{K-1} I(x_k;\bsm{y}_k,\bsm{y}_{k+1}|x_{k-1}). \label{opt.chain}
 \end{align} 
 Then, $\mathcal{P}_3$ is reformulated as 
\begin{align}
    \mathcal{P}_4 \quad \max_{\bsm{\theta}} \quad& \sum_{k=1}^{K-1} I(x_k;\bsm{y}_k,\bsm{y}_{k+1}|x_{k-1}) \notag \\ 
    \textrm{s.t.}\quad& \eqref{opt.obj}. 
\end{align}

$\mathcal{P}_4 $ can be similarly solved by following the AO procedures described in the above subsection. Specifically, we first recast $\mathcal{P}_4 $ based the WMMSE method as 
\begin{align}
    \mathcal{P}_5 \quad \max_{\bsm{\theta}, \{\Sigma_k\},\{\bsm{\phi}_k\}} \quad& \sum_{k=1}^{K-1} f_k(\bsm{\theta}, \Sigma_k,\bsm{\phi}_k) \notag \\ 
    \textrm{s.t.}\quad& \eqref{opt.obj}. 
\end{align}
where $f_k(\bsm{\theta}, \Sigma_k,\bsm{\phi}_k) \triangleq  \log(\Sigma_k) - \Sigma_k \mbs{E} \left( x_k - \bsm{\phi}_k^{\rm H}\left[\begin{array}{c}
    \bsm{y}_k \\ \bsm{y}_{k+1}
    \end{array}\right] \right)^2 $ and $\bsm{\phi}_k \in \mathbb{C}^{2M}$ and $\Sigma_k\in \mathbb{R}$ being the auxiliary variables. Then $\bsm{\theta}$ and $(\{\Sigma_k\},\{\bsm{\phi}_k\})$ are alternatively optimized as follows.

$\emph{1)}$ For given $\bsm{\theta}$, the optimized $\{\Sigma_k\})$ and $(\{\bsm{\phi}_k\}$ are given as 
\begin{align} 
    \Sigma_k = \left(1 - \bsm{\mathcal{C}}_{x\bsm{y}}^k (\bsm{\mathcal{C}}_{\bsm{y}\bsm{y}}^k)^{-1} \bsm{\mathcal{C}}_{\bsm{y}x}^k \right)^{-1}  \quad \text{and} \quad \bsm{\phi}_k = (\bsm{\mathcal{C}}_{\bsm{y}\bsm{y}}^k)^{-1} \bsm{\mathcal{C}}_{\bsm{y}x}^k, \forall k, \label{opt.phi1}
\end{align}
where $\bsm{\mathcal{C}}_{\bsm{y}x}^k = P\left[\begin{array}{c} \bsm{h}_k^{\rm UB}+ \frac{1}{2}\bsm{H}_k\bsm{\theta} \\ \frac{1}{2}\tilde{\bsm{H}}_{k+1}\bsm{\theta}  \end{array}\right] $,  
$ \bsm{\mathcal{C}}_{\bsm{y}\bsm{y}}^k = P\left[\begin{array}{c} \bsm{h}_k^{\rm UB} + \frac{1}{2}\bsm{H}_k\bsm{\theta} \\ \frac{1}{2}\tilde{\bsm{H}}_{k+1}\bsm{\theta}  \end{array}\right] \left[\begin{array}{c} \bsm{h}_k^{\rm UB}\bsm{\theta} + \frac{1}{2} \bsm{H}_k\bsm{\theta} \\ \frac{1}{2}\tilde{\bsm{H}}_{k+1}  \end{array}\right]^{\rm H} \\+ \left[\begin{array}{cc} \bsm{\mathcal{Q}}^{/\bsm{\theta}}_k & \mathbf{0}  \\ \mathbf{0} & \tilde{\bsm{\mathcal{Q}}}^{/\bsm{\theta}}_k \end{array}\right]  $ with $\tilde{\bsm{\mathcal{Q}}}^{/\bsm{\theta}}_k = \bsm{\mathcal{Q}}^{/\bsm{\theta}}_{k+1} + P \left(\bsm{h}_{k+1}^{\rm UB} + \frac{1}{2}\bsm{H}_{k+1}\bsm{\theta} \right)\left(\bsm{h}_{k+1}^{\rm UB} + \frac{1}{2}\bsm{H}_{k+1}\bsm{\theta}\right)^{\rm H} $, and $\bsm{\mathcal{C}}_{x\bsm{y}}^k = \left(\bsm{\mathcal{C}}_{\bsm{y}x}^k\right)^{\rm H} $.  
    
$\emph{2)}$ For given $(\{\Sigma_k\},\{\bsm{\phi}_k\})$, the optimization of $\mathcal{P}_5$ over $\bsm{\theta}$ is 
\begin{align}
    \mathcal{P}_{5.1} \quad \min_{\bsm{\theta}} \quad& \bsm{\theta}^{\rm H} \bsm{\Lambda}'  \bsm{\theta} 
    + 2\Re\left\{  (\bsm{\alpha}')^{\rm H} \bsm{\theta} \right\}  \notag \\ 
    \textrm{s.t.}\quad& \eqref{opt.obj}. 
\end{align}
where $\bsm{\Lambda}' = \sum_{k=1}^{K-1} \bsm{\Lambda}_k$, $\bsm{\alpha}' =\sum_{k=1}^{K-1}\bsm{\alpha}_k$,
\begin{align}
    \bsm{\Lambda}_k = \Sigma_k\left[\begin{array}{c} \bsm{H}_k \\ \tilde{\bsm{H}}_{k+1}  \end{array}\right]^{\rm H} \bsm{\phi}_k \bsm{\phi}_k^{\rm H} \left[\begin{array}{c} \bsm{H}_k \\ \tilde{\bsm{H}}_{k+1}  \end{array}\right] + \Sigma_k\left[\begin{array}{c} \mathbf{0} \\ \bsm{H}_{k+1}  \end{array}\right]^{\rm H} \bsm{\phi}_k \bsm{\phi}_k^{\rm H} \left[\begin{array}{c} \mathbf{0} \\ \bsm{H}_{k+1}  \end{array}\right], \label{opt.Lambda1}
\end{align}  
\begin{align}
    \bsm{\alpha}_k = \Sigma_k \left(\left[\begin{array}{c} 2\bsm{h}_k^{\rm UB} \\ \mathbf{0}  \end{array}\right]^{\rm H} \bsm{\phi}_k - 1 \right) \bsm{\phi}_k^{\rm H} \left[\begin{array}{c} \bsm{H}_k \\ \tilde{\bsm{H}}_{k+1}  \end{array}\right], \forall k. \label{opt.alpha1}
\end{align}
$\mathcal{P}_{5.1} $ is solved by the MM algorithm similarly as $\mathcal{P}_{3.2}$. 

\begin{algorithm}[t]
    \caption{The RAO Algorithm }
    \label{alg_RAO}
\begin{algorithmic}[1]
    \REQUIRE  $\bsm{h}^{\rm UB}$, $\bsm{H}$, $\tilde{\bsm{H}}$, and $\bsm{\mathcal{Q}}$
    \STATE Initialize $\bsm{\theta}$ randomly
    \REPEAT
    \STATE $\forall k:$ Update $(\Sigma_k, \bsm{\phi}_k)$ according to \eqref{opt.phi1}
    \STATE $\forall k:$ Update $(\bsm{\Lambda}_k, \bsm{\alpha}_k)$ according to \eqref{opt.Lambda1} and \eqref{opt.alpha1}, respectively
    \STATE Update $\bsm{\theta}$ by the MM algorithm given in Algorithm \ref{alg:MM}
    \UNTIL convergence criterion is met  
    \ENSURE $\bsm{\theta}$
\end{algorithmic}  
\end{algorithm}

The overall RAO algorithm is summarized in Algorithm \ref{alg_RAO}. Compared with the AO algorithm, the computational complexity in Line $3$ is reduced to $\mathcal{O}(KM^3)$, $\mathcal{O}(K^2)$ times smaller than that of the former.


\section{Bilinear Detection Design}

The main difficulty of the receiver design in the FRM-OFDM system lies in the joint detection of the user symbols $\bsm{x}$ and the RIS data $\bsm{c}$. To tackle this bilinear recovery problem, we design a message passing based algorithm under the Bayesian framework to alternatively detect these two sources of information.

\begin{figure}[t]
    \centering
    \includegraphics[width=2.3 in]{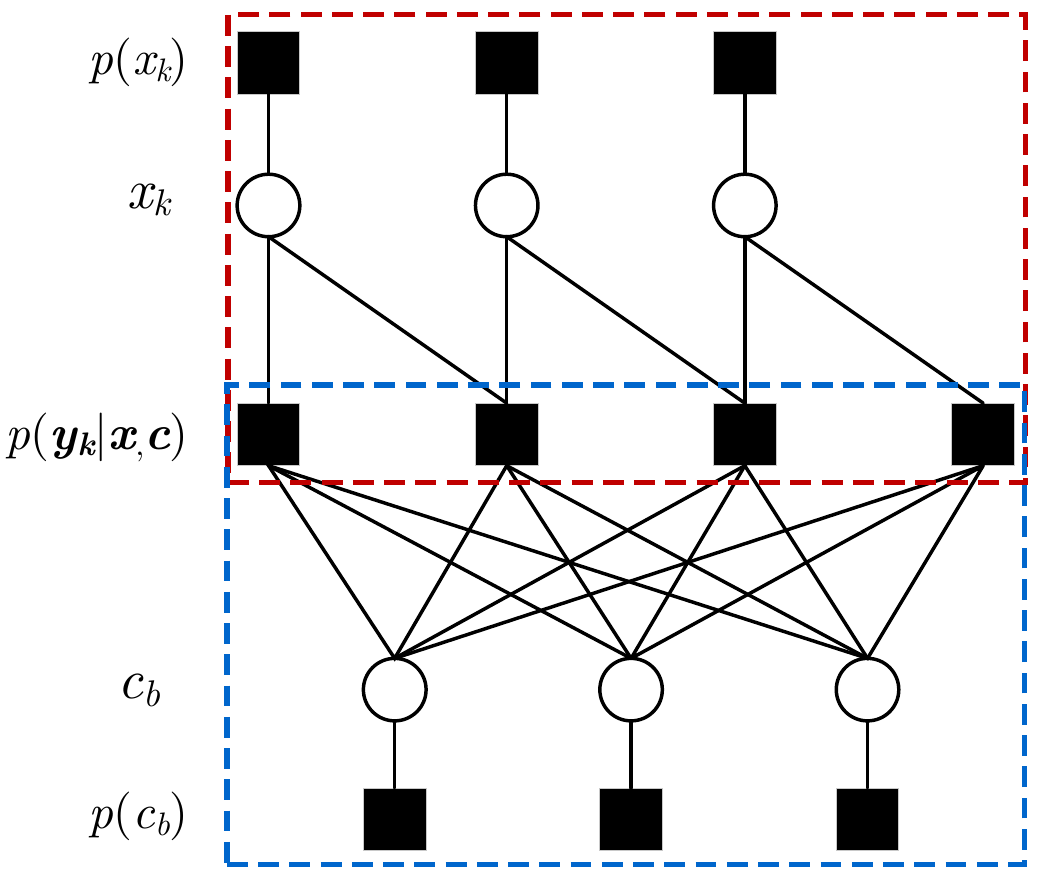}
    \caption{The factor graph representation for the joint probability in \eqref{pro_1} with $K=4$ and $B=3$.}\label{factor_graph}
\end{figure}

Motivated by the maximum \emph{a posteriori} principle, we formulate the recovery problem as  
\begin{align}
    \left( \hat{\bsm{x}},\hat{\bsm{c}}\right) = \arg\max_{\bsm{x},\bsm{c}} p(\bsm{x},\bsm{c}|\bsm{y}),
\end{align}
where $p(\bsm{x},\bsm{c}|\bsm{y}) $ is the joint posterior probability of $\bsm{x} $ and $\bsm{c}$, given by
\begin{subequations} \label{pro_1}
    \begin{align}
    p(\bsm{x},\bsm{c}|\bsm{y}) &\varpropto p(\bsm{y}|\bsm{x},\bsm{c}) p(\bsm{x}) p(\bsm{c}) \label{pro_1.a} \\
    & = \left[ \prod_{k=1}^K  p(\bsm{y}_{k}|\bsm{x},\bsm{c} )\right]  \left[ \prod_{k=1}^{K-1}  p(x_{k})\right]\left[\prod_{b=1}^{\rm B} p(c_b) \right], \label{pro_1.b} 
\end{align}
\end{subequations}
where \eqref{pro_1.a} employs Bayes’ rule and \eqref{pro_1.b} resorts to the independency of the entries of $\bsm{x}$ and $\bsm{c}$.
A factor graph to depict the joint probability in \eqref{pro_1} is given in Fig.~\ref{factor_graph}, where a hollow circle represents a ``variable node'' and a solid square represents a ``factor node''. 
Next, we derive our detection algorithm by sum-product message passing over the factor graph. 

\subsection{The Detection of $\bsm{x}$}

We start from detecting $\bsm{x}$ by message passing over the upper part of Fig.~\ref{factor_graph}. 
For ease of exposition, some auxiliary variables are introduced, namely, $\bsm{\xi}_k \triangleq \bsm{h}_k^{\rm UB} + \bsm{H}_k \bsm{\Theta}\bsm{s} = [\xi_{k,1}, \cdots, \xi_{k,M}]^{\rm T}$ and $\tilde{\bsm{\xi}}_k \triangleq \tilde{\bsm{H}}_{k} \bsm{\Theta}(\textbf{1}_{N}-\bsm{s}) = [\tilde{\xi}_{k,1}, \cdots, \tilde{\xi}_{k,M}]^{\rm T}$, $\forall k$. Recall that $\bsm{s} = \bsm{T}\bsm{c}$ and $\bsm{T} \triangleq \textbf{I}_B \otimes \textbf{1}_L$. For large $B\rightarrow \infty$, the CLT motivates the treatment of $\{\xi_{k,m}\}$ and $\{\tilde{\xi}_{k,m}\}$ as circular symmetric complex Gaussian (CSCG) variables, i.e., $\xi_{k,m} \sim \mathcal{CN}(\xi_{k,m}; \mu_{\xi_{k,m}}, \nu_{\xi_{k,m}} )$ and $\tilde{\xi}_{k,m} \sim \mathcal{CN}(\tilde{\xi}_{k,m}; \mu_{\tilde{\xi}_{k,m}}, \nu_{\tilde{\xi}_{k,m}})$, $\forall m,k$.
Then, we have 
\begin{align} 
    \bsm{y}_{k} &= x_k \bsm{\xi}_k + x_{k-1} \tilde{\bsm{\xi}}_k+ \bsm{w}_k, \forall k.  \label{Markov}
\end{align}  
Eq.~\eqref{Markov} shows that $\bsm{y}_{k} \to x_k \to \bsm{y}_{k+1}$ forms a Markov chain. Thus, the related messages passing reduced to forward and backward recursions. The forward message recursion calculates the messages passed from $\bsm{y}_k$ to $x_k$, while the backward recursion calculates the messages from $\bsm{y}_{k+1}$ to $x_k$, $\forall k$. The detailed message passing process is described below.

\subsubsection{Forward message recursion}

The message from $\bsm{y}_k$ to $x_k$ is 
\begin{align}
\Delta_{\bsm{y}_k \rightarrow x_k}(x_k) &\varpropto \sum_{x_{k-1}, \bsm{c}} p(\bsm{y}_k|x_{k},x_{k-1},\bsm{c}) \Delta_{x_{k-1}\rightarrow \bsm{y}_k }(x_{k-1}) \prod_{b=1}^B \Delta_{c_b \rightarrow \bsm{y}_k}(c_b)  \notag\\
&\approx  \sum_{x_{k-1}} \int_{\bsm{\xi}_k,\tilde{\bsm{\xi}}_k} \mathcal{CN}(\bsm{y}_k; x_k \bsm{\xi}_k \!+\! x_{k-1} \tilde{\bsm{\xi}}_k, \sigma^2 \mathbf{I}_M) \Delta_{x_{k-1}\rightarrow \bsm{y}_k }(x_{k-1}) \notag\\
&~~\prod_{m=1}^M \mathcal{CN}(\xi_{k,m}; \mu_{\xi_{k,m}}, \nu_{\xi_{k,m}}) \mathcal{CN}(\tilde{\xi}_{k,m}; \mu_{\tilde{\xi}_{k,m}}, \nu_{\tilde{\xi}_{k,m}}), \forall k, \label{MP_1}
\end{align}
where \eqref{MP_1} is based on the fact that $\bsm{c} \to (\bsm{\xi}_k, \tilde{\bsm{\xi}}_k) \to \bsm{y}_k$ forms a Markov chain and the approximation that $\{\xi_{k,m}\}$ and $\{ \tilde{\xi}_{k,m}\}$ are independent of each other. With the message from $\bsm{c}$, i.e., $\prod_{b=1}^B \Delta_{c_b \rightarrow \bsm{y}_k}(c_b) $, we have $\mu_{\xi_{k,m}} = h_{k,m}^{\rm UB} + \sum_{n=1}^{N}h_{k,m,n} \theta_n\hat{s}_n$, $\nu_{\xi_{k,m}} = \sum_{n=1}^{N}h_{k,m,n}^2 \nu_{s_n}$, $\mu_{\tilde{\xi}_{k,m}} = \sum_{n=1}^{N}\tilde{h}_{k,m,n} \theta_n(1- \hat{s}_n)$, and $\nu_{\tilde{\xi}_{k,m}} = \sum_{n=1}^{N}\tilde{h}_{k,m,n}^2 \nu_{s_n}$, $\forall k, m$, where $\hat{s}_n = \hat{c}_b$,  $\nu_{s_n} = \nu_{c_b}$ if $\lceil \frac{n}{L}\rceil = b$, $\forall n$, and $(\{\hat{c}_b\}, \{\nu_{c_b}\})$ are respectively the estimated means and variances of $\{c_b\}$ in the last iteration. $\Delta_{x_{k-1} \rightarrow \bsm{y}_k} (x_{k-1})$ in \eqref{MP_1} is the message from $x_{k-1}$ to $\bsm{y}_k$, given by
\begin{align}
 \Delta_{x_{k-1} \rightarrow \bsm{y}_k} (x_{k-1}) \varpropto p(x_{k-1})\Delta_{\bsm{y}_{k-1} \rightarrow x_{k-1}}(x_{k-1}), \forall k,
\end{align}
where $\Delta_{\bsm{y}_0 \rightarrow x_0}(x_{0}) \triangleq 1$.

\begin{figure}[t] 
    \centering
    \includegraphics[width=2.5 in]{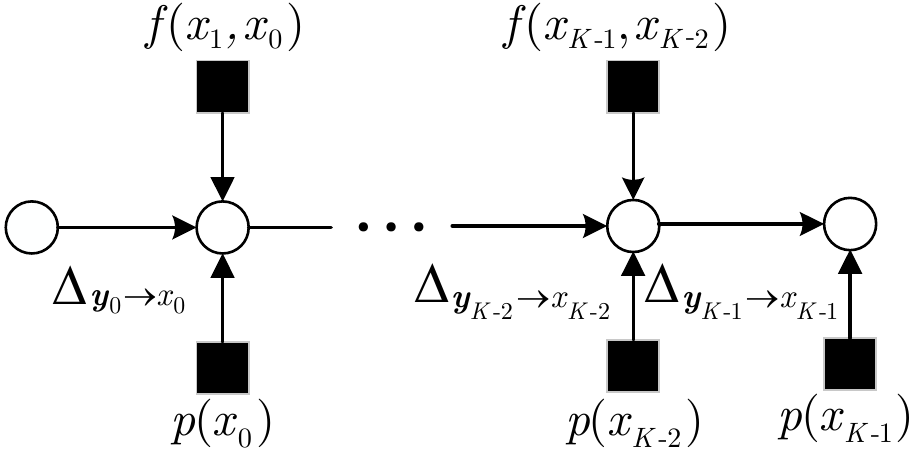}
    \caption{Forward recursion for the message computation where $f(x_k, x_{k-1})$ is defined in eq.~\eqref{MP_2}.}\label{ForRec}
\end{figure}

Eq.~\eqref{MP_1} involves integration over all the $2M$ variables in $\bsm{\xi}_k$ and $\tilde{\bsm{\xi}}_k$. To reduce the computational complexity, we apply the MRC technique to $\bsm{y}_k$ as 
\begin{align} 
    \frac{\bsm{\mu}_{\bsm{\xi}_k}^{\rm H}\bsm{y}_{k}}{\|\bsm{\mu}_{\bsm{\xi}_k}\|^2}  &= x_{k}\frac{\bsm{\mu}_{\bsm{\xi}_k}^{\rm H} \bsm{\xi}_k}{\|\bsm{\mu}_{\bsm{\xi}_k}\|^2} + x_{k-1}\frac{\bsm{\mu}_{\bsm{\xi}_k}^{\rm H} \tilde{\bsm{\xi}}_k}{\|\bsm{\mu}_{\bsm{\xi}_k}\|^2}  + \frac{\bsm{\mu}_{\bsm{\xi}_k}^{\rm H}\bsm{w}_k}{\|\bsm{\mu}_{\bsm{\xi}_k}\|^2} ,
\end{align}
where $\bsm{\mu}_{\bsm{\xi}_k} = [\mu_{\xi_{k,1}}, \cdots, \mu_{\xi_{k,M}}]^{\rm T}$. Denote by $y_k^{\rm F} = \frac{\bsm{\mu}_{\bsm{\xi}_k}^{\rm H}\bsm{y}_{k}}{\|\bsm{\mu}_{\bsm{\xi}_k}\|^2} $,  $h_k^{\rm F} = \frac{\bsm{\mu}_{\bsm{\xi}_k}^{\rm H} \bsm{\xi}_k}{\|\bsm{\mu}_{\bsm{\xi}_k}\|^2} \sim \mathcal{CN}\left(\cdot; \mu_{h_k^{\rm F}}, \nu_{h_k^{\rm F}}\right)$, where $\mu_{h_k^{\rm F}} = 1$ and $\nu_{h_k^{\rm F}} = \frac{\sum_{m=1}^M \mu_{\xi_{k,m}}^2 \nu_{\xi_{k,m}}}{(\|\bsm{\mu}_{\bsm{\xi}_k}\|^2)^2} $, $\tilde{h}_k^{\rm F} = \frac{\bsm{\mu}_{\bsm{\xi}_k}^{\rm H} \tilde{\bsm{\xi}}_k}{\|\bsm{\mu}_{\bsm{\xi}_k}\|^2} \sim \mathcal{CN}\left(\cdot;  \mu_{\tilde{h}_k^{\rm F}}, \nu_{\tilde{h}_k^{\rm F}}\right)$, where $\mu_{\tilde{h}_k^{\rm F}} = \frac{\bsm{\mu}_{\bsm{\xi}_k}^{\rm H} \bsm{\mu}_{\tilde{\bsm{\xi}}_k}}{\|\bsm{\mu}_{\bsm{\xi}_k}\|^2}$ and $ \nu_{\tilde{h}_k^{\rm F}} = \frac{\sum_{m=1}^M \mu_{\xi_{k,m}}^2 \nu_{\tilde{\xi}_{k,m}}}{(\|\bsm{\mu}_{\bsm{\xi}_k}\|^2)^2} $, and $\frac{\bsm{\mu}_{\bsm{\xi}_k}^{\rm H}\bsm{w}_k}{\|\bsm{\mu}_{\bsm{\xi}_k}\|^2} \sim \mathcal{CN}(\cdot ;0,\sigma^2)$.
Then, \eqref{MP_1} is simplified to  
\begin{align}
    &\Delta_{\bsm{y}_k \rightarrow x_k} (x_k)  \notag\\
    &\varpropto \sum_{x_{k-1}} \int_{h_k^{\rm F}, \tilde{h}_k^{\rm F}} \mathcal{CN} \left(y_k^{\rm F}; x_kh_k^{\rm F} + x_{k-1}\tilde{h}_k^{\rm F}, \sigma^2 \right) \Delta_{x_{k-1}\rightarrow \bsm{y}_k}(x_{k-1}) \mathcal{CN}(h_k^{\rm F}; 1, \nu_{h_k^{\rm F}}) \mathcal{CN}(\tilde{h}_k^{\rm F}; \mu_{\tilde{h}_k^{\rm F}}, \nu_{\tilde{h}_k^{\rm F}}) \notag\\
    &=  \sum_{x_{k-1}} p(x_{k-1}) \Delta_{\bsm{y}_{k-1} \rightarrow x_{k-1}}(x_{k-1}) \underbrace{\mathcal{CN}\left(y_k^{\rm F}; x_k + x_{k-1}\mu_{\tilde{h}_k^{\rm F}}, \sigma^2 + x_k^2 \nu_{h_k^{\rm F}} + x_{k-1}^2 \nu_{\tilde{h}_k^{\rm F}}\right)}_{\triangleq f(x_k, x_{k-1})}, \forall k.  \label{MP_2}
\end{align}
\eqref{MP_2} implies a forward message recursion over $\Delta_{\bsm{y}_k \rightarrow x_k} (x_k), \forall k$, which is represented by Fig.~\ref{ForRec}.

\begin{figure}[t] 
    \centering
    \includegraphics[width=2.5 in]{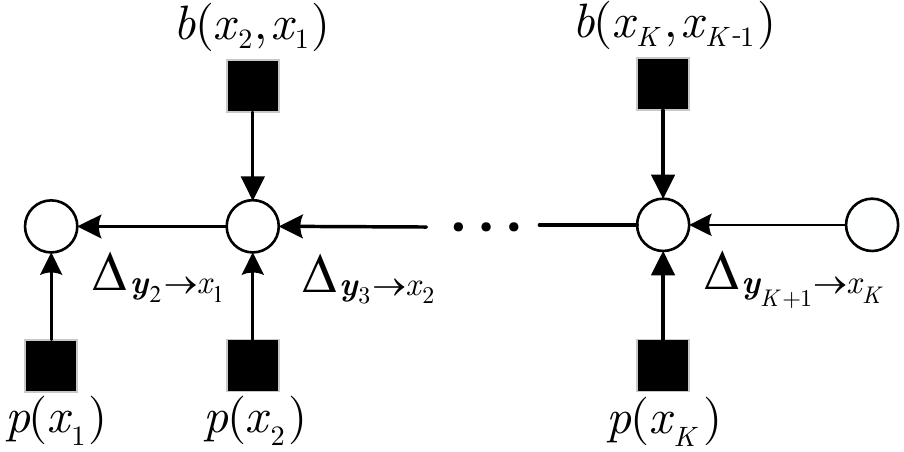}
    \caption{Backward recursion for the message computation where $b(x_k, x_{k-1})$ is defined in eq.~\eqref{MP_4}.}\label{BackRec} 
\end{figure} 

\subsubsection{Backward message recursion}
The message from $\bsm{y}_{k}$ to $x_{k-1}$ is 
\begin{align}
&\Delta_{\bsm{y}_{k} \rightarrow x_{k-1}} (x_{k-1}) \varpropto \sum_{x_{k}, \bsm{c}} p(\bsm{y}_{k}|x_{k},x_{k-1},\bsm{c}) \Delta_{x_{k}\rightarrow \bsm{y}_{k} }(x_k) \prod_{b=1}^B \Delta_{c_b \rightarrow \bsm{y}_k}(c_b), \forall k,  \label{MP_3}
\end{align}
where $\Delta_{x_{k} \rightarrow \bsm{y}_{k}} (x_{k})$ is the message from $x_{k}$ to $\bsm{y}_{k}$ given as
\begin{align}
\Delta_{x_k \rightarrow \bsm{y}_k} (x_k) \varpropto p(x_k)\Delta_{\bsm{y}_{k+1} \rightarrow x_k}(x_k), \forall k,
\end{align}
with $\Delta_{\bsm{y}_{K+1} \rightarrow x_K}(x_K) \triangleq 1$.

Similar to the forward message recursion, we apply the backward MRC to $\bsm{y}_{k}$ for ease the computation of \eqref{MP_3}. The backward MRC is designed as 
\begin{align} 
    \frac{ \bsm{\mu}_{\tilde{\bsm{\xi}}_k}^{\rm H}\bsm{y}_{k}}{\|\bsm{\mu}_{\tilde{\bsm{\xi}}_k}\|^2} &= x_k \frac{\bsm{\mu}_{\tilde{\bsm{\xi}}_k}^{\rm H} \bsm{\xi}_k}{\|\bsm{\mu}_{\tilde{\bsm{\xi}}_k}\|^2}  + x_{k-1} \frac{\bsm{\mu}_{\tilde{\bsm{\xi}}_k}^{\rm H} \tilde{\bsm{\xi}}_k}{\|\bsm{\mu}_{\tilde{\bsm{\xi}}_k}\|^2} + \frac{\bsm{\mu}_{\tilde{\bsm{\xi}}_k}^{\rm H}\bsm{w}_k}{\|\bsm{\mu}_{\tilde{\bsm{\xi}}_k}\|^2},
\end{align}
where $\bsm{\mu}_{\tilde{\bsm{\xi}}_k} = [\mu_{\tilde{\xi}_{k,1}}, \cdots, \mu_{\tilde{\xi}_{k,M}}]^{\rm T}$. Denote by $y_k^{\rm B} = \frac{\bsm{\mu}_{\tilde{\bsm{\xi}}_k}^{\rm H}\bsm{y}_{k}}{\|\bsm{\mu}_{\tilde{\bsm{\xi}}_k}\|^2} $,  $h_k^{\rm B} = \frac{\bsm{\mu}_{\tilde{\bsm{\xi}}_k}^{\rm H} \bsm{\xi}_k}{\|\bsm{\mu}_{\tilde{\bsm{\xi}}_k}\|^2} \sim \mathcal{CN}\left(\cdot; \mu_{h_k^{\rm B}}, \nu_{h_k^{\rm B}}\right)$, where $\mu_{h_k^{\rm B}} = \frac{\bsm{\mu}_{\tilde{\bsm{\xi}}_k}^{\rm H} \bsm{\mu}_{\bsm{\xi}_k}}{\|\bsm{\mu}_{\tilde{\bsm{\xi}}_k}\|^2}$ and $\nu_{h_k^{\rm B}} = \frac{\sum_{m=1}^M \mu_{\tilde{\xi}_{k,m}}^2 \nu_{\xi_{k,m}}}{(\|\bsm{\mu}_{\tilde{\bsm{\xi}}_k}\|^2)^2} $, $\tilde{h}_k^{\rm B} = \frac{\bsm{\mu}_{\tilde{\bsm{\xi}}_k}^{\rm H} \tilde{\bsm{\xi}}_k}{\|\bsm{\mu}_{\tilde{\bsm{\xi}}_k}\|^2} \sim \mathcal{CN}\left(\cdot;  \mu_{\tilde{h}_k^{\rm B}}, \nu_{\tilde{h}_k^{\rm B}}\right)$, where $\mu_{\tilde{h}_k^{\rm B}} = 1$ and $ \nu_{\tilde{h}_k^{\rm B}} = \frac{\sum_{m=1}^M \mu_{\tilde{\xi}_{k,m}}^2 \nu_{\tilde{\xi}_{k,m}}}{(\|\bsm{\mu}_{\tilde{\bsm{\xi}}_k}\|^2)^2} $, and $\frac{\bsm{\mu}_{\tilde{\bsm{\xi}}_k}^{\rm H}\bsm{w}_k}{\|\bsm{\mu}_{\tilde{\bsm{\xi}}_k}\|^2} \sim \mathcal{CN}(\cdot ;0,\sigma^2)$.
Then, \eqref{MP_3} is simplified to  
\begin{align}
    &\Delta_{\bsm{y}_{k} \rightarrow x_{k-1}} (x_{k-1})\notag\\
    &\varpropto \sum_{x_{k}} p(x_{k}) \Delta_{\bsm{y}_{k+1} \rightarrow x_{k}}(x_K) \underbrace{\mathcal{CN}\left(y_k^{\rm B}; x_k\mu_{h_k^{\rm B}} + x_{k-1}, \sigma^2 + x_k^2 \nu_{h_k^{\rm B}} + x_{k-1}^2 \nu_{\tilde{h}_k^{\rm B}}\right)}_{\triangleq  b(x_k, x_{k-1})}, \forall k.  \label{MP_4}
\end{align}  
The backward message recursion over $\Delta_{\bsm{y}_{k} \rightarrow x_{k-1}}(x_{k-1}), \forall k$ in \eqref{MP_4} is represented by Fig.~\ref{BackRec}.
The estimated mean and variance of $x_k$ are respectively computed as 
\begin{align}
    \hat{x}_k &= \sum_{x_k} x_k p(x_k) \Delta_{\bsm{y}_k\rightarrow x_k}(x_k)\Delta_{\bsm{y}_{k+1} \rightarrow x_k}(x_k),\\
    \nu_{x_k} &= \sum_{x_k} |x_k - \hat{x}_k|^2 p(x_k) \Delta_{\bsm{y}_k\rightarrow x_k}(x_k)\Delta_{\bsm{y}_{k+1} \rightarrow x_k}(x_k), \forall k.
\end{align}

\subsection{Message Passing for Detecting $\bsm{c}$}

The detection of $\bsm{c}$ is carried out by message passing over the lower part of Fig.~\ref{factor_graph}. First, we calculate the message passed from its upper part. 
For computational convenience, each $x_k$ is approximated as a Gaussian variable with probability $\mathcal{CN}(; \hat{x}_k, \nu_{x_k})$, $\forall k\in \mathcal{I}_{k-1}$ during the lower part message passing process. 
The residual of $x_k$ is defined as $x_k - \hat{x}_k \sim \mathcal{CN}(; 0, \nu_{x_k})$. To proceed, we recast \eqref{SysMod_5} as 
\begin{align} 
    \bsm{y}_{k} - \hat{x}_k\bsm{h}_{k}^{\rm UB} - \hat{x}_k\tilde{\bsm{H}}_k\bsm{\theta} & = \hat{x}_k (\bsm{H}_k - \tilde{\bsm{H}}_k)\bsm{\Theta}\bsm{T} \bsm{c}  + \bsm{n}_k, \forall k,  \label{MP_c}
\end{align}
where $\bsm{n}_k$ is the equivalent noise for detecting $\bsm{c}$, defined as
\begin{align}
    \bsm{n}_k= (x_k - \hat{x}_k)\left( (\bsm{H}_k - \tilde{\bsm{H}}_k )\bsm{\Theta}\bsm{T} \bsm{c} + \bsm{h}_{k}^{\rm UB} +\tilde{\bsm{H}}_k\bsm{\theta} \right)+ \bsm{w}_k, \forall k. 
\end{align}
Eq.~\eqref{MP_c} provides an explicit linear model for the estimation of $\bsm{c}$ with observations $\{\bsm{y}_{k} - \hat{x}_k\bsm{h}_{k}^{\rm UB} - \hat{x}_k\tilde{\bsm{H}}_k\bsm{\theta}\}$ and measurements $\{\hat{x}_k (\bsm{H}_k - \tilde{\bsm{H}}_k)\bsm{\Theta}\bsm{T} \}$.  An existing message passing algorithm to handle this model is the GAMP \cite{rangan2011generalized}, which is used here to recover $\bsm{c}$. The details of the GAMP algorithm is omitted here for brevity.

\subsection{Algorithm Summary}

\begin{algorithm}[!htbp]
    \vspace{0.2mm}
    \caption{\textbf{\!:} \vspace{0.15mm} The BMP algorithm}
    \label{Alg_MessPass}
\begin{algorithmic}[1]
    \REQUIRE $\bsm{y}$, $\{\bsm{h}_k^{\rm UB}\}$, $\{\bsm{H}_k\}$, $\{\tilde{\bsm{H}}_k\}$, $\mathcal{X}$  \\
    \hspace{-0.7cm}Initialization: $\hat{c}_b = \frac{1}{2}$, 
    $\nu_{c_b} = \frac{1}{4}$, $\forall b$ \\
    \hspace{-0.8cm} \textbf{repeat}   \\
    \STATE\hspace{0.3cm}$\forall k, m:$ $\mu_{\xi_{k,m}} = h_{k,m}^{\rm UB} + \sum_{n=1}^{N}h_{k,m,n} \theta_n\hat{s}_n$, $\nu_{\xi_{k,m}} = \sum_{n=1}^{N}h_{k,m,n}^2 \nu_{s_n}$ \\
    \hspace{1.5cm} $\mu_{\tilde{\xi}_{k,m}} = \sum_{n=1}^{N}\tilde{h}_{k,m,n} \theta_n(1- \hat{s}_n)$, $\nu_{\tilde{\xi}_{k,m}} = \sum_{n=1}^{N}\tilde{h}_{k,m,n}^2 \nu_{s_n}$ \\  
    \STATE\hspace{0.3cm}$\forall k:$ $\mu_{h_k^{\rm F}} = 1$, $\nu_{h_k^{\rm F}} = \frac{\sum_{m=1}^M \mu_{\xi_{k,m}}^2 \nu_{\xi_{k,m}}}{(\|\bsm{\mu}_{\bsm{\xi}_k}\|^2)^2} $, $\mu_{\tilde{h}_k^{\rm F}} = \frac{\bsm{\mu}_{\bsm{\xi}_k}^{\rm H} \bsm{\mu}_{\tilde{\bsm{\xi}}_k}}{\|\bsm{\mu}_{\bsm{\xi}_k}\|^2}$, $ \nu_{\tilde{h}_k^{\rm F}} = \frac{\sum_{m=1}^M \mu_{\xi_{k,m}}^2 \nu_{\tilde{\xi}_{k,m}}}{(\|\bsm{\mu}_{\bsm{\xi}_k}\|^2)^2} $ \\
    \hspace{1cm} $\mu_{h_k^{\rm B}} = \frac{\bsm{\mu}_{\tilde{\bsm{\xi}}_k}^{\rm H} \bsm{\mu}_{\bsm{\xi}_k}}{\|\bsm{\mu}_{\tilde{\bsm{\xi}}_k}\|^2}$, $\nu_{h_k^{\rm B}} = \frac{\sum_{m=1}^M \mu_{\tilde{\xi}_{k,m}}^2 \nu_{\xi_{k,m}}}{(\|\bsm{\mu}_{\tilde{\bsm{\xi}}_k}\|^2)^2} $, $\mu_{\tilde{h}_k^{\rm B}} = 1$, $ \nu_{\tilde{h}_k^{\rm B}} = \frac{\sum_{m=1}^M \mu_{\tilde{\xi}_{k,m}}^2 \nu_{\tilde{\xi}_{k,m}}}{(\|\bsm{\mu}_{\tilde{\bsm{\xi}}_k}\|^2)^2} $ \\ 
    \hspace{0.3cm}\textbf{for} $k = 1,2,\ldots, K-1$  \quad \% Forward recursion \\
    \STATE\hspace{1cm}$\Delta_{\bsm{y}_k \rightarrow x_k} (x_k) \varpropto \sum_{x_{k-1}} p(x_{k-1}) \Delta_{\bsm{y}_{k-1} \rightarrow x_{k-1}}(x_{k-1}) f(x_k, x_{k-1})$ \\
    \hspace{0.3cm}\textbf{end for}\\
    \hspace{0.3cm}\textbf{for} $k = K,K-1,\ldots, 2$  \quad \% Backward recursion \\
    \STATE\hspace{1cm}$\Delta_{\bsm{y}_k \rightarrow x_{k-1}} (x_{k-1}) \varpropto \sum_{x_k} p(x_k) \Delta_{\bsm{y}_{k+1} \rightarrow x_k}(x_k) b(x_k, x_{k-1})$ \\
    \hspace{0.3cm}\textbf{end for}\\
    \STATE\hspace{0.3cm}$\forall k:$ $\hat{x}_k = \sum_{x_k} x_k p(x_k) \Delta_{\bsm{y}_k\rightarrow x_k}(x_k)\Delta_{\bsm{y}_{k+1} \rightarrow x_k}(x_k)$ \\
    \hspace{1cm} $\nu_{x_k} = \sum_{x_k} |x_k - \hat{x}_k|^2 p(x_k) \Delta_{\bsm{y}_k\rightarrow x_k}(x_k)\Delta_{\bsm{y}_{k+1} \rightarrow x_k}(x_k)$ \\
    \STATE\hspace{0.3cm}$\forall k:$ Detecting $\bsm{c}$ based on the linear model given in eq.~\eqref{MP_c} by GAMP algorithm.\\
    \hspace{-0.8cm} \textbf{until} a certain stopping criterion is met \\
    \ENSURE $\{\hat{x}_k\}$ and $\{\hat{c}_b\}$.  
\end{algorithmic} 
\end{algorithm}

The overall BMP algorithm is summarized in Algorithm~\ref{Alg_MessPass}. 
In specific, Line $1$ updates the means and variances of the auxiliary variables $\{\xi_{k,m}\}$ and $\{\tilde{\xi}_{k,m}\}$. Line $2$ updates the means and variances of the auxiliary variables $\{h_k^{\rm F}\}$, $\{\tilde{h}_k^{\rm F}\}$, $\{h_k^{\rm B}\}$, and $\{\tilde{h}_k^{\rm B}\}$. Line $3$ recursively computes the  messages from $\{\bsm{y}_k\}$ to $\{x_k\}$, where $\Delta_{\bsm{y}_0 \rightarrow x_0} (x_0) =1 $ and $p(x_0) = \delta(0)$. 
Line $4$ recursively computes the messages from $\{\bsm{y}_k\}$ to $\{x_{k-1}\}$, where $\Delta_{\bsm{y}_{K+1} \rightarrow x_K} (x_K) =1 $ and $p(x_k) = \delta(0)$.
Line $5$ updates of the estimated means $\{\hat{x}_k\}$ and variances $\{\nu_{{x}_k}\}$ for $\{x_k\}$. 
Line $6$ performs the GAMP algorithm to detect $\bsm{c}$.  
The stopping criterion is defined as
$\mbs{E} \left| \left| y_{k,m} - \hat{x}_k \mu_{\xi_{k,m}} \!-\! \hat{x}_{k-1}\mu_{\tilde{\xi}_{k,m}} \right|^2 - \sigma^2 \right| \leq \epsilon$, where $\epsilon$ is the tolerance parameter, or is when the iteration number reaching to the preset maximum value.

We now analyze the computational complexity of the proposed BMP algorithm.
The complexity in Line $1$ is $\mathcal{O}(KMN)$. The complexity in Line $2$ is $\mathcal{O}(KM)$. 
The complexities in Line $3$ and Line $4$ are both $\mathcal{O}(|\mathcal{X}|^2)$, and thus the complexities of the ``for'' loops therein are $\mathcal{O}(K|\mathcal{X}|^2)$.
The complexity in Line $5$ is $\mathcal{O}(K|\mathcal{X}|)$. The complexity in Line $6$ depends on the complexity of the GAMP algorithm, which is  $\mathcal{O}(KMB)$ \cite{rangan2011generalized}. 

\section{Numerical Results} \label{sec.Simul}
 
This section conducts numerical experiments to evaluate the performance of the proposed FRM-OFDM scheme for RIS-aided transmission, the AO and RAO algorithms for the system optimization, and the BMP algorithm for the bilinear signal detection. First, we describe the simulation setup.

\subsection{Simulation Setup} \label{sec.sub.Setting}

\begin{figure}[t] 
    \centering
    \includegraphics[width=2.3 in]{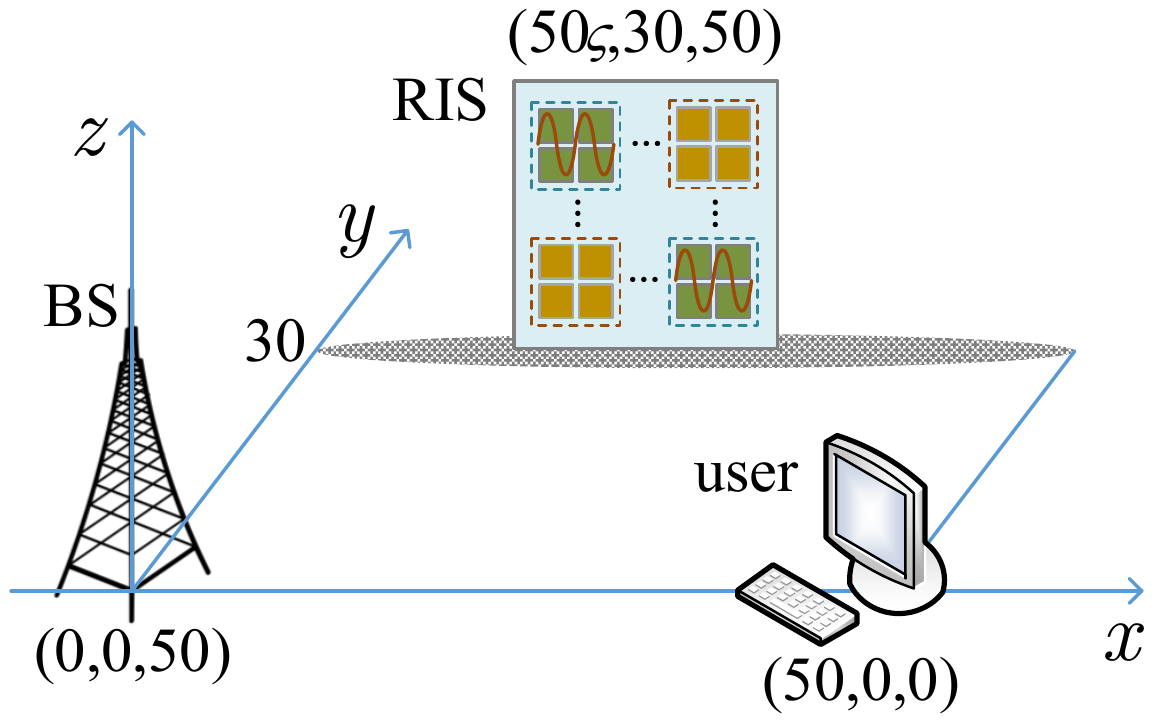}
    \caption{The simulated RIS-aided FRM-OFDM scenario.}\label{location} 
\end{figure}
We consider a three-dimensional (3D) Cartesian coordinate system shown in Fig.~\ref{location}. The BS is placed at $(0; 0; 50)$, the user is located at $(50; 0; 0)$, and the RIS is randomly deployed at  $(50 \varsigma ; 30; 50)$ with $ \varsigma$ is a random number between $[0,1]$. The multipath numbers of the user-BS, user-RIS, and RIS-BS links are set to $8$, $8$, and $6$, respectively.
The CP length is selected to perfectly suppress the ISI. Usually, the deployment location of the RIS is chosen to favor line-of-sight (LoS) transmission between the RIS and the BS/user. Thus, the first paths of the user-RIS and RIS-BS links are modeled as LoS paths, and the remaining paths are treated as non-LoS (NLoS) paths characterized by Rayleigh fading. For the user-BS link, the LoS path is assumed to be blocked by obstacles, with only NLoS paths.
The entries of the NLoS paths are independently taken from the CSCG distribution $\mathcal{CN}(0, 1)$. Since the size of the RIS array is practically much smaller than the distance between the RIS and the BS/user, a far-field scenario is adopted.
The RIS is assumed to be equipped with a two-dimensional ($2$D) uniform rectangular array, with dimension $L_x \times L_y$, where $L_x=8$ and $L_y= \frac{N}{8}$. Then, the LoS path can be expressed as the $2$D array steering vector, given by
\begin{align}
\mathbf{a}(\vartheta,\psi)&=\mathbf{a}_{\textsf{az}}(\vartheta,\psi)\otimes \mathbf{a}_{\textsf{el}}(\vartheta,\psi),
\end{align}
where $\vartheta$ is the azimuth angle, and $\psi$ is the elevation angle.
$\mathbf{a}_{\textsf{az}}(\vartheta,\psi) \in \mathbb{C}^{L_x \times 1}$ and $\mathbf{a}_{\textsf{el}}(\vartheta,\psi)\in \mathbb{C}^{L_y \times 1}$ are the uniform linear array (ULA) steering vectors given as
\begin{align}
\hspace{-0.2cm}\left[\mathbf{a}_{\textsf{az}}(\vartheta,\psi)\right]_n &= e^{-j \frac{2\pi d_x (n-1)}{\varrho}\cos(\vartheta) \sin(\psi)}, \forall n \in \mathcal{I}_{L_x}, \\
 \left[\mathbf{a}_{\textsf{el}}(\vartheta,\psi)\right]_n &= e^{j \frac{2\pi d_y (n-1)}{\varrho}\sin(\vartheta) \sin(\psi)}, \forall n \in \mathcal{I}_{L_y},
\end{align}
where the wavelength of propagation $\varrho$ is set to $ 3\times10^8/f_c$ meters, with $f_c = 915$ MHz being the carrier frequency, $d_x=d_y=3\times10^7/f_c$ meters are respectively the horizontal and vertical sizes of a single RIS element.
$\kappa_1 = 3$ dB and $\kappa_2 = 10$ dB denote the power ratios of the LoS component to the NLoS component for the user-RIS link and for the RIS-BS link, respectively.
The free-space path loss model proposed in \cite{tang2020Path} is adopted for the cascaded user-RIS-BS channel, given by 
\begin{align}
    PL = G_{\rm user}G_{\rm RIS}G_{\rm BS} \frac{L_x L_y d_x d_y(3\times10^8/f_c)^2}{64 \pi^3 d_{\rm UR}d_{\rm RB}},
\end{align}
where $G_{\rm user}=0$ dBi, $G_{\rm RIS}=5$ dBi, and  $G_{\rm BS}=0$ dBi are respectively the antenna gain at the user, the RIS, and the BS, and
$d_{\rm RB}/d_{\rm UR}$ is the distance between the RIS and the BS/user. The free-space path loss of the user-BS link is modeled as $\beta = \beta_0 d^{-\alpha }$, where $\beta_0=-30$ dB, $d$ is the link distance, and $\alpha=2.2$.

In simulations,  the elements of $\bsm{X}$ are randomly taken from the quadrature phase shift keying (QPSK) modulation with Gray-mapping. The noise variance $\sigma^2$ is set to $-60$ dBw.
 The iteration numbers are set to $100$ for the MM algorithm and $200$ for the AO and RAO algorithms. The maximum iteration number is set to $10$ for the BMP algorithm. 
 The error tolerance threshold for the BMP algorithm is set to  $\epsilon = 10^{-3}$. 
The presented simulation results are obtained by taking average over at least $1000$ random realizations. The unit of the rate is bit per channel use per SC (bpcu). The number of RIS blocks $B$ is set to its maximum, i.e., $N$, in Figs.~$6$, $7$, $8$, and $9$.

\subsection{Simulations of RIS Phase shift Optimization Design} 
 
\begin{figure}[!htbp] 
    \centering
    \includegraphics[width=3 in]{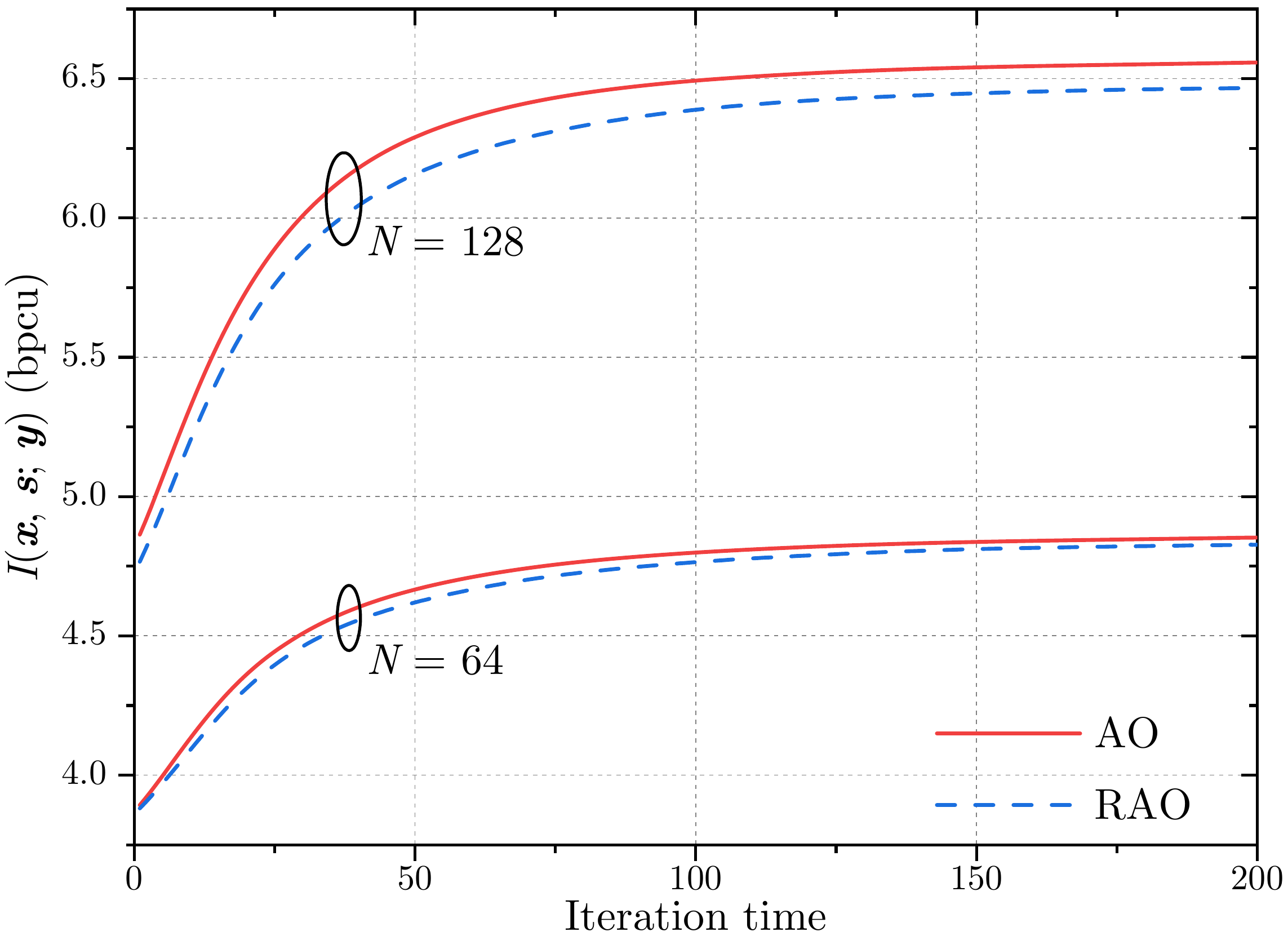}
    \caption{Comparison of $I(\bsm{x},\bsm{s};\bsm{y})$ versus the iteration time with the AO and RAO algorithms.} \label{Fig_Opt_Converge}
\end{figure} 
\begin{figure}[!htbp] 
    \centering
    \includegraphics[width=3 in]{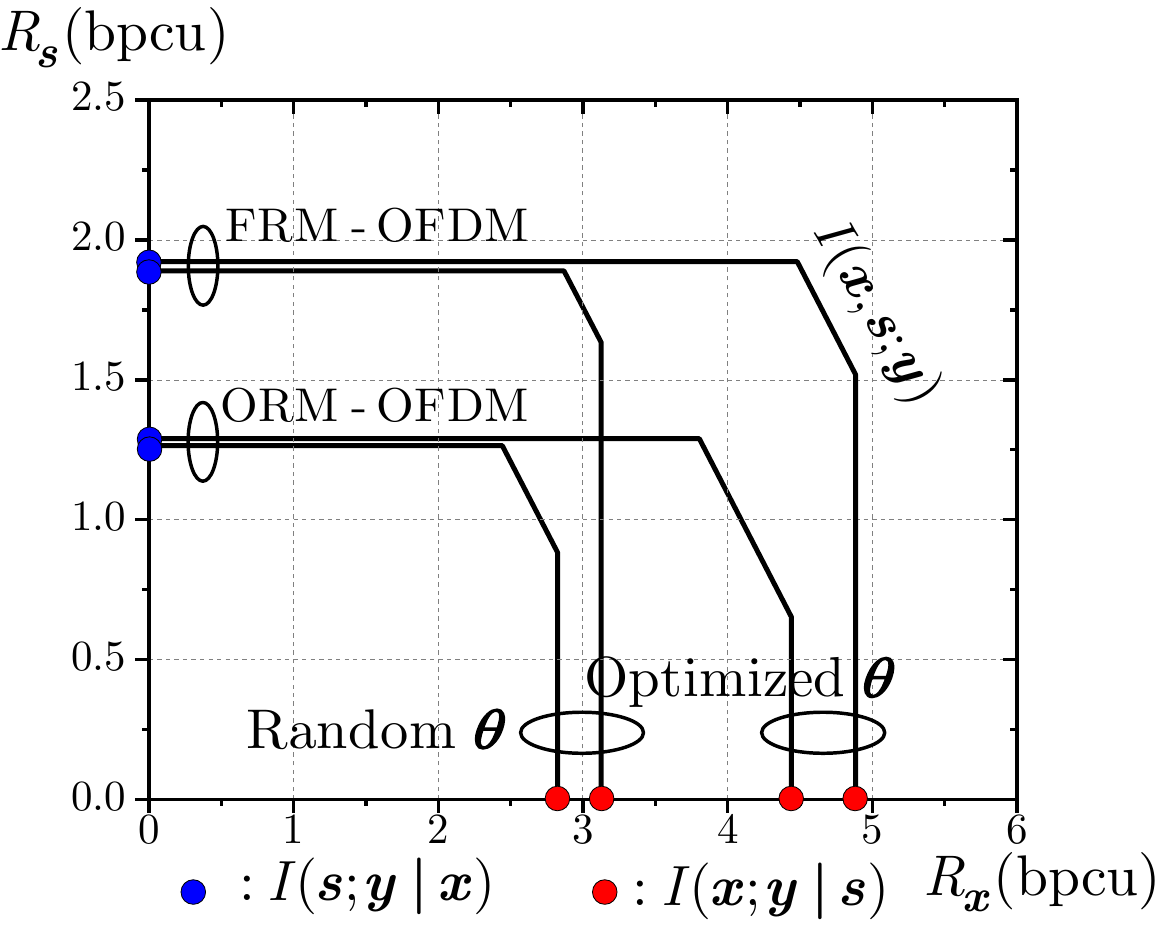}
    \caption{Rate region comparison of the ORM-OFDM scheme and the FRM-OFDM scheme with $M=8$, $K=16$, $N=128$, and $P=0$ dBw.} \label{Fig_Opt_Capa}
\end{figure} 
\begin{figure}[!htbp] 
    \centering
    \includegraphics[width=3.5 in]{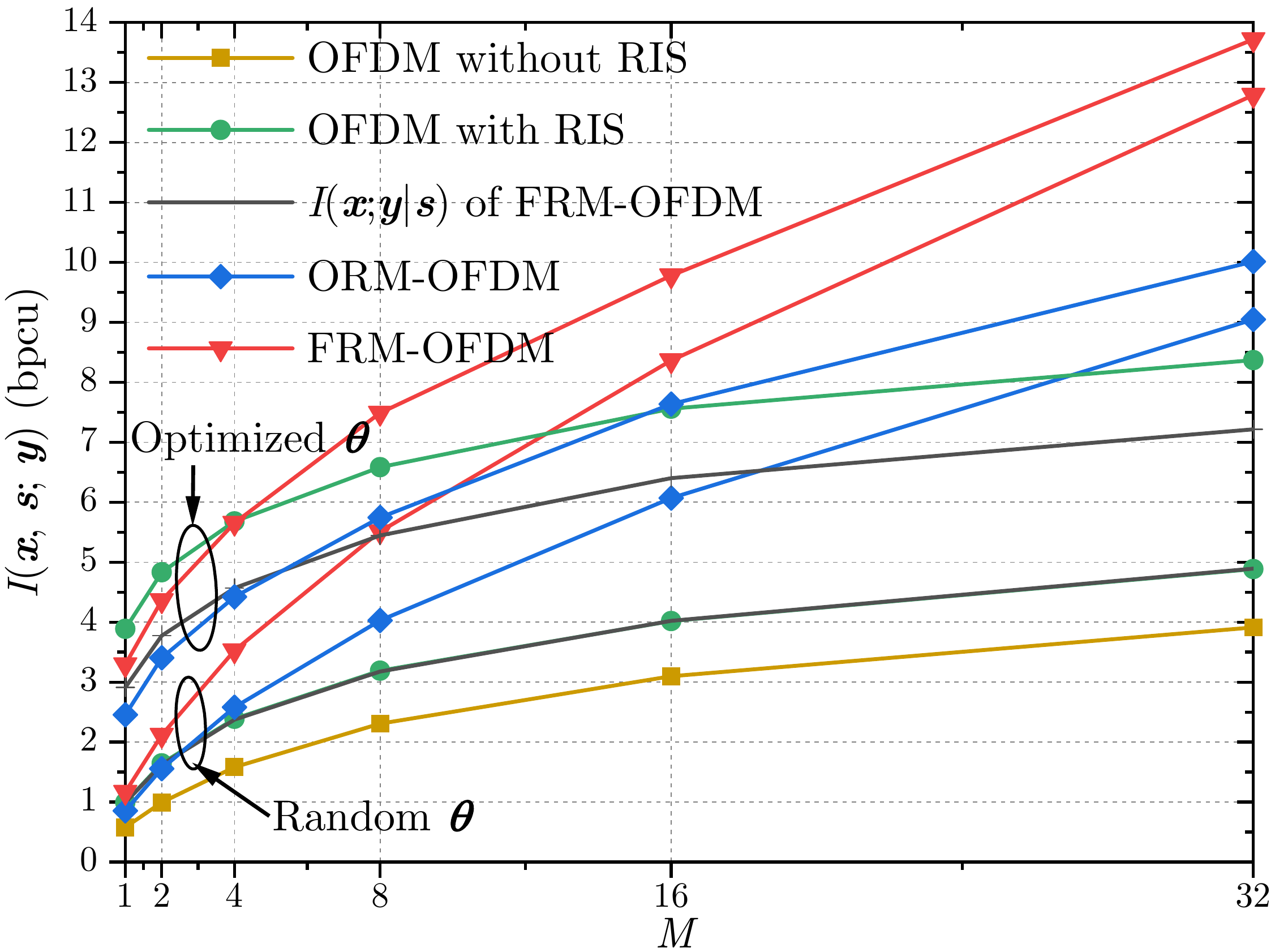}
    \caption{Sum rate comparison versus the number of receiver antennas $M$ with $K=8$, $N=192$, and $P=0$ dBw.} \label{Fig_Opt_M}
\end{figure}

Fig.~\ref{Fig_Opt_Converge} compares the sum rate performance of the AO and RAO algorithms versus the iteration time. The parameter settings are: $M=8$, $K=16$, and $P=0$ dBw. We see that for both $N=64$ and $N=128$, the RAO algorithm shows marginal rate performance degradation as compared to the AO algorithm. Thus, in the following figures, we only present the optimization results by the RAO algorithm for a lower computational burden.  

\begin{figure}[!htbp] 
    \centering
    \includegraphics[width=3.5 in]{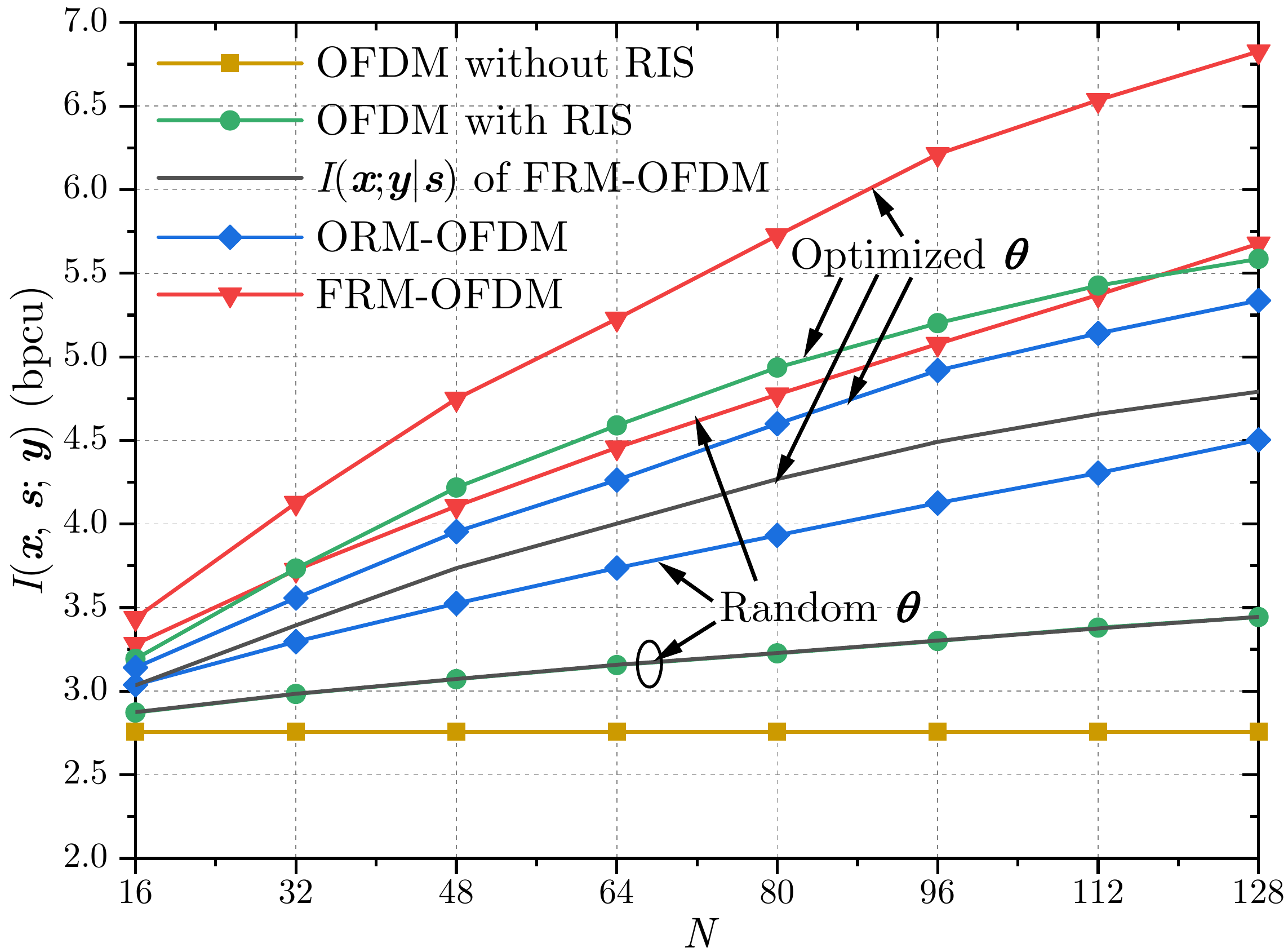}
    \caption{Sum rate comparison versus the number of RIS elements $N$ with $K=8$, $M=12$, and $P=0$ dBw.} \label{Fig_Opt_N}
\end{figure} 
 
Fig.~\ref{Fig_Opt_Capa} compares the rate region of the proposed FRM-OFDM scheme with the existing ORM-OFDM scheme. The parameter settings are: $M=8$, $K=16$, $N=128$, and $P=0$ dBw. We see that no matter $\bsm{\theta}$ is optimized or randomly chosen from $[0, 2\pi]$, the ORM-OFDM scheme covers a wider rate region than the ORM-OFDM scheme. Specially, for both the $\bsm{\theta}$ cases, the FRM-OFDM scheme outperforms the ORM-OFDM scheme by about $1.2$ bpcu in terms of RIS rate, i.e., $I(\bsm{s}, \bsm{y}|\bsm{x})$, and by about $0.5$ bpcu in terms of user rate, i.e., $I(\bsm{x}, \bsm{y}|\bsm{s})$. We also see that, by optimization, both the FRM-OFDM scheme and the ORM-OFDM scheme achieve about $1.5$ bpcu gain in user rate, but have marginal improvement on the RIS rate.

Fig.~\ref{Fig_Opt_M} compares the sum rate of the FRM-OFDM scheme with the ORM-OFDM scheme, the RIS-aided OFDM system, and the OFDM without RIS case versus the number of receive antennas $M$. For comparison, we also provide the user rate $I(\bsm{x};\bsm{y}|\bsm{s})$ of the FRM-OFDM system.  The parameter settings are: $K=8$, $N=192$, and $P=0$ dBw. 
 We see that, before optimization, the FRM-OFDM outperforms the ORM-OFDM, the RIS-aided OFDM, and the OFDM without RIS by about $4$, $8$, and $9$ bpcu, respectively, at $M=32$. At the same $M$ setting, by optimization, the rate gains of the FRM-OFDM, the ORM-OFDM, and the RIS-aided OFDM are about $1$, $1$, and $3$ bpcu, respectively. 
 The user rate $I(\bsm{x};\bsm{y}|\bsm{s})$ of the FRM-OFDM system tightly approaches the rate of the OFDM system for random $\bsm{\theta}$ case, while lower than the latter by about $1$ bpcu after optimization. We conjecture that this is because the FRM-OFDM system sacrifices some RIS passive beamforming capability in exchanging for RIS information transfer. Fig.~\ref{Fig_Opt_M} also shows that, the FRM-OFDM has a sharper slope in rate growth than the RIS-aided OFDM as the increase of $M$, which demonstrates the multiplexing gain obtained by RIS information transfer.

Fig.~\ref{Fig_Opt_N} shows the sum rate comparison of the same systems as Fig.~\ref{Fig_Opt_M} versus the number of RIS elements $N$. The parameter settings are: $K=8$, $M=12$, and $P=0$ dBw. 
We see that before optimization, the FRM-OFDM outperforms the ORM-OFDM from about $0.3$ bpcu to about $1$ bpcu, and outperforms the RIS-aided OFDM from about $0.5$ bpcu to about $2$ bpcu, as $N$ ranging from $32$ to $256$. By optimization, the rate gaps between the FRM-OFDM and the ORM-OFDM, and between the FRM-OFDM and the RIS-aided OFDM are about $1.5$ and $1.3$ bpcu, respectively, at $N=128$. Comparing the curves of the ``OFDM without RIS'' and the ``FRM-OFDM'', we see that the rate gain by the assistance of an RIS can be as high as $4$ bpcu when $N=128$.

\subsection{Simulations of Bilinear Receiver Design} 
\begin{figure}[!htbp] 
    \centering
    \includegraphics[width=3.5 in]{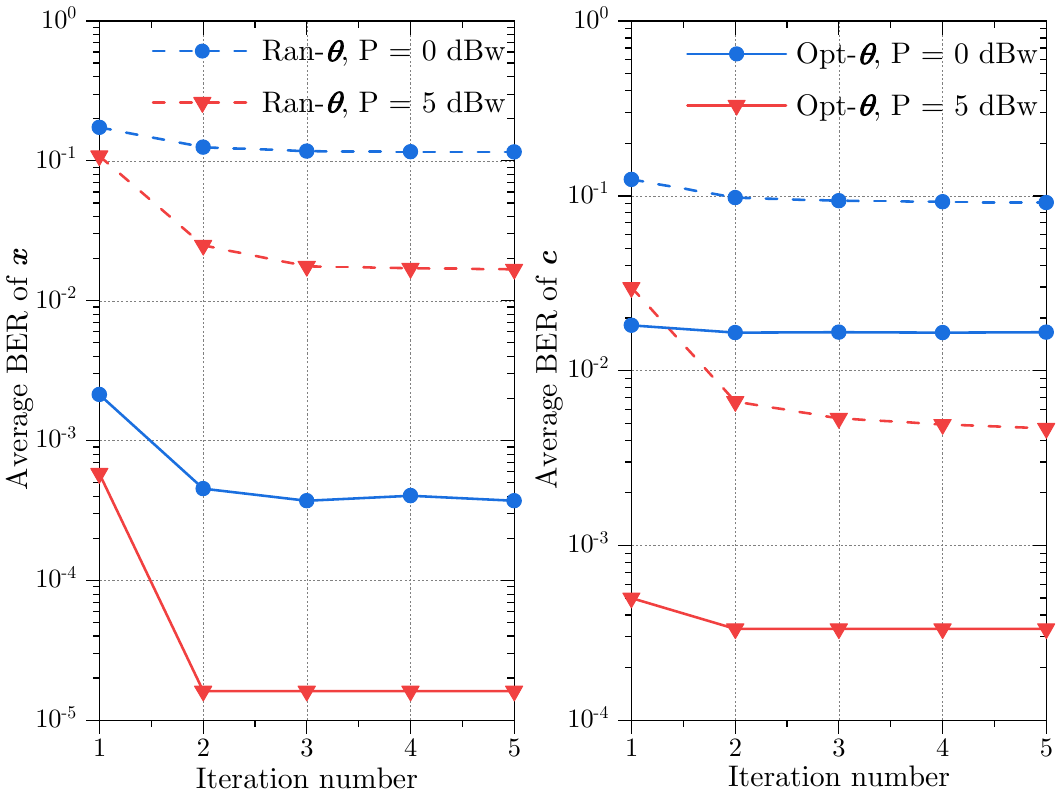}
    \caption{Convergence behavior of the TMP algorithm with $M=8$, $K=32$, $B=20$, and $L=4$.} \label{Det_Conv}
\end{figure} 
\begin{figure}[!htbp] 
    \centering
    \includegraphics[width=3.5 in]{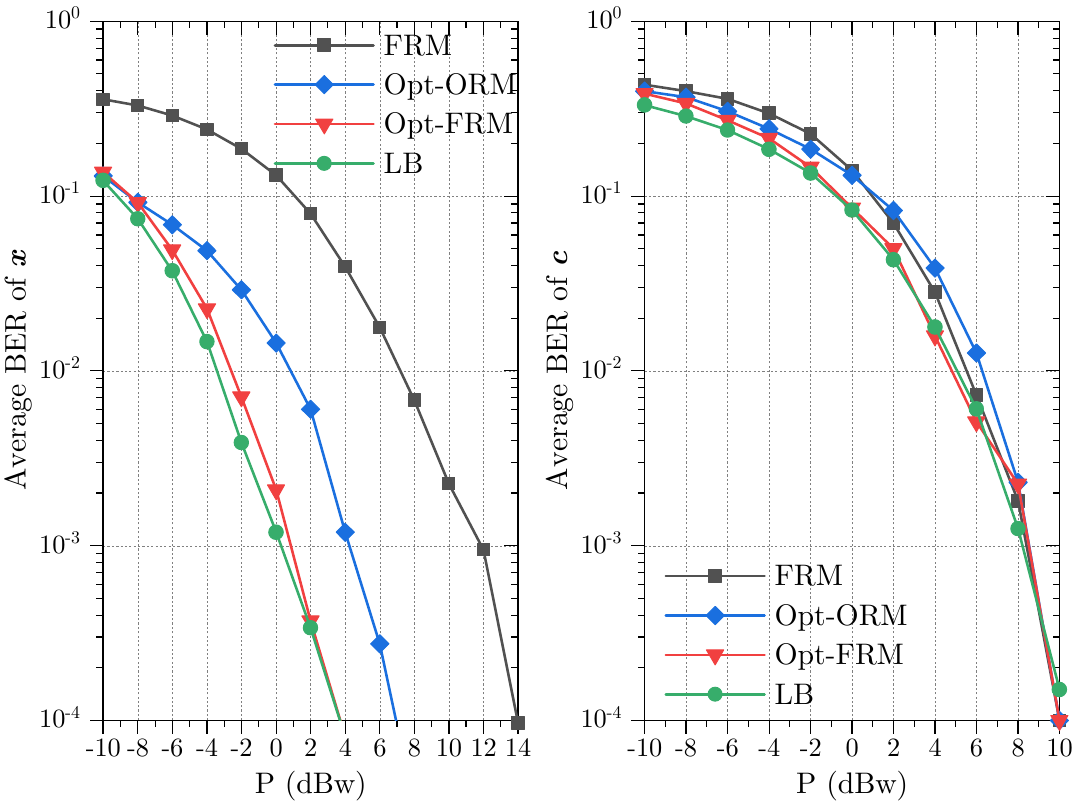} 
    \caption{ Average BERs of $\bsm{x}$ and $\bsm{c}$ versus the transmitted power $P$ with $M=8$, $K=32$, $B=20$, and $L=4$.} \label{Fig_Det_SNR}
\end{figure}
\begin{figure}[!htbp] 
    \centering
    \includegraphics[width=3.5 in]{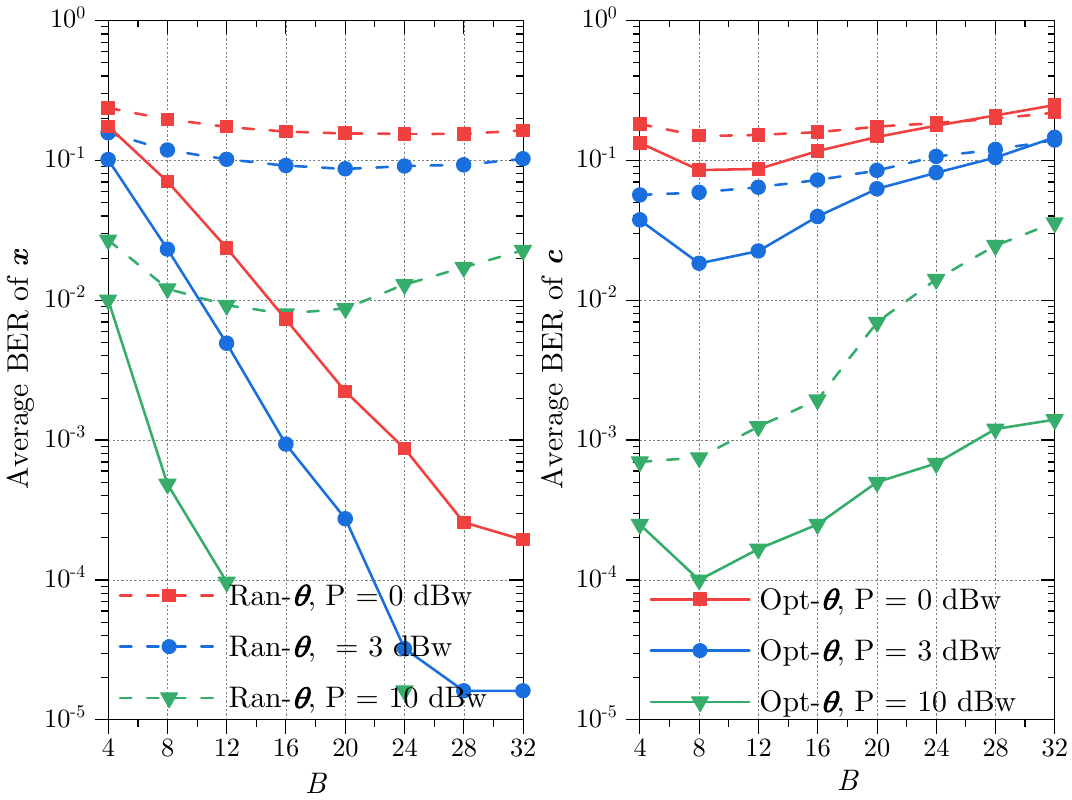}
    \caption{ Average BERs of $\bsm{x}$ and $\bsm{c}$ versus the RIS information blocks $B$ with $M=8$, $K=32$ and $L=4$.}  \label{Fig_Det_B}
\end{figure}

The convergence behavior of the TMP algorithm is presented in Fig~\ref{Det_Conv}. The parameter settings are: $M=8$, $K=32$, $B=20$, and $L=4$. The ``Ran-$\bsm{\theta}$'' represents the random $\bsm{\theta}$ case and the ``Opt-$\bsm{\theta}$'' represents the optimized $\bsm{\theta}$ case. We consider $P=0$ dBw and $P=5$ dBw for comparison. We see that in all the considered cases, the TMP algorithm converges within $4$ times of iteration. 

Fig.~\ref{Fig_Det_SNR} shows the average bit error rate (BER) performance of the BMP algorithm versus the transmitted power $P$. We consider the ``FRM'' (the FRM-OFDM system with random $\bsm{\theta}$), the ``Opt-ORM'' (the ORM-OFDM system with optimized $\bsm{\theta}$), the ``Opt-FRM'' (the FRM-OFDM system with optimized $\bsm{\theta}$), and the lower bound (LB) (which is obtained by retrieving $\bsm{x}$ (or $\bsm{c}$) with perfectly known $\bsm{c}$ (or $\bsm{x}$) in the FRM-OFDM system). The parameter settings are: $M=8$, $K=32$ and $B=20$ and $L=4$. We see that the FRM-OFDM tightly approaches the lower bound when detecting both $\bsm{x}$ and $\bsm{c}$. 
When detecting $\bsm{x}$, the Opt-FRM outperforms the Opt-ORM by about $3$ dBw, and outperforms the FRM by about $10$ dBw at the average BER of $\bsm{x}=10^{-4}$. The average BER of $\bsm{c}$ is insensitive to the considered cases.

Fig.~\ref{Fig_Det_B} shows the average BER performance of the BMP algorithm versus the number of RIS blocks $B$. The parameter settings are: $M=8$, $K=32$ and $L=4$.
We present three comparison pairs of $P=0$, $3$, and $10$ dBw.
We see that in general, the average BER of $\bsm{x}$ drops with the increase of $B$. For $\bsm{c}$, the average BER drops from $B=4$ to $B=8$ but then rises. 
This is because more RIS information block increases the reflected signal power, and thus enhances the detection of $\bsm{x}$. In return, the better BER performance of $\bsm{x}$ also helps the detection of $\bsm{c}$. However, the larger $B$ also puts more detection difficulty on $\bsm{c}$, and as $B$ becomes relatively too large, the BER of $\bsm{c}$ starts to increase. We also see that the BER performance of $\bsm{x}$ is nonsensitive to the BER of $\bsm{c}$ thanks to the strong direct link. 


\section{Conclusions}

This paper studied the joint PB and PIT for the RIS-aided systems. We proposed an FRM-OFDM scheme to modulate the incident EM waves by random frequency-hopping at different RIS elements. We showed that, with a Gaussian approximation on the distribution of the received signals, the MMAC channel of the FRM-OFDM system can be equivalent to a MIMO channel in the sense of RIS phase shift optimization, and then proposed AO algorithm to maximize the sum rate of the system. A low-complexity RAO algorithm is further developed to avoid the direct matrix inversion over the MIMO channel with negligible performance degradation. Moreover, we designed a BMP algorithm to effectively retrieve both the user symbols and the RIS data. In both the rate region and the signal recovery, the proposed FRM-OFDM scheme showed better performance than the existing ORM-OFDM scheme.

\bibliographystyle{IEEEtran}
\bibliography{FRM-OFDM}

\end{document}